\documentclass[%
 reprint,
 aps,
 nofootinbib,
]{revtex4-2}

\preprint{APS/123-QED}

\usepackage{mathrsfs,mathpazo,amsmath,mathtools,physics}

\usepackage{graphicx}
\usepackage{dcolumn}
\usepackage{hyperref}
\usepackage{xcolor}
\usepackage{ulem}
\usepackage{tikz}
\usepackage{listings}
\usepackage{lipsum, babel}
\usepackage{multirow}
\usepackage[font=small,labelfont=bf,justification=justified]{subcaption}
\usepackage[font=small,labelfont=bf,justification=justified]{caption}

\usepackage[makeroom]{cancel}
\usepackage{float}
\usepackage{pgf}

\tikzset{
    level/.style = {
        ultra thick,
        black,
    },
    connect/.style = {
        dashed,
        red
    }
}
\usetikzlibrary{backgrounds,calc}
\usepackage[compat=1.1.0]{tikz-feynman}
\usetikzlibrary{shapes.misc}
\usepackage{multirow}
\allowdisplaybreaks
\makeatletter
\newcommand*\rel@kern[1]{\kern#1\dimexpr\macc@kerna}
\newcommand*\widebar[1]{%
  \begingroup
  \def\mathaccent##1##2{%
    \rel@kern{0.8}%
    \overline{\rel@kern{-0.8}\macc@nucleus\rel@kern{0.2}}%
    \rel@kern{-0.2}%
  }%
  \macc@depth\@ne
  \let\math@bgroup\@empty \let\math@egroup\macc@set@skewchar
  \mathsurround\z@ \frozen@everymath{\mathgroup\macc@group\relax}%
  \macc@set@skewchar\relax
  \let\mathaccentV\macc@nested@a
  \macc@nested@a\relax111{#1}%
  \endgroup
}

\begin{document}

\preprint{APS/123-QED}

\title{Modified Urca neutrino emissivity at finite temperature}

\author{Lami Suleiman}
\email{lami.suleiman@obspm.fr}
 \affiliation{Laboratoire Univers et Th\'{e}ories, Observatoire de Paris, Universit\'{e} PSL, Université Paris Cit\'{e}, CNRS, F-92190 Meudon, France}
\affiliation{Nicolaus Copernicus Astronomical Center of the Polish Academy of Sciences, ul. Bartycka 18, 00-716 Warszawa, Poland 
}
\thanks{Present address: Nicholas and Lee Begovich Center for Gravitational Wave Physics and Astronomy, California State University Fullerton, Fullerton, California 92831, USA.}
\def\ls{\textcolor{purple}}

\author{Micaela Oertel}%
 \email{micaela.oertel@obspm.fr}
\affiliation{Laboratoire Univers et Th\'{e}ories, Observatoire de Paris, Universit\'{e} PSL, Université Paris Cit\'{e}, CNRS, F-92190 Meudon, France
}
\def\mo{\textcolor{blue}}

\author{Marco Mancini}
\email{marco.mancini@obspm.fr}
\affiliation{IDP, UMR 7013 - CNRS - Univ. Orléans- Univ. Tours.
Université d'Orléans, rue de Chartres, BP 6759,  45067 Orléans Cedex 2, France}
\affiliation{Laboratoire Univers et Th\'{e}ories, Observatoire de Paris, Universit\'{e} PSL, Université Paris Cit\'{e}, CNRS, F-92190 Meudon, France}
\def\mm{\textcolor{green}}

\date{\today}

\begin{abstract}
\begin{description}

\item[Background] 
Charged-current neutrino-nucleon reactions, generally called Urca processes, are crucial actors of a neutron star's thermal evolution. The so-called direct processes show a pronounced threshold under which the reaction is kinematically suppressed. This suppression does not apply to "modified" Urca processes, which involve interaction with an additional nucleon. Calculations of the modified Urca neutrino rates were established for cold neutron star matter and for dilute hot matter, in both cases under strong assumptions. 
\item[Purpose]
In this paper, we revise the calculations of the modified Urca neutrino rates for dense and hot matter, and for different compositions. We study the influence of different approximations used in previous computations.  
\item[Methods]
We derive expressions for the rates of modified Urca neutrino emissivity within thermal field theory and perform the phase space integration numerically using mainly importance sampling Monte Carlo techniques. The neutrino emissivity of modified and direct processes are established and compared. 
\item[Results]
We find in particular that the modified Urca process is not necessarily suppressed with respect to the direct process above the threshold of the latter at moderate densities and temperatures, in contrast to what is generally assumed. Numerical results are confirmed by an estimation of the ratio of modified to direct Urca rates with a simple analytic approximation, thereby showing the regimes of suppression for the modified processes depending on temperature and density. 
\item[Conclusion] 
These results show that modified Urca rates have to be considered carefully upon evaluating neutrino opacities in dense and warm matter.
\end{description}
\end{abstract}

\maketitle


\section{Introduction}

Neutron stars (NSs) are compact stars containing matter at the highest densities in the Universe, exceeding the density at the center of atomic nuclei. In addition, neutron star matter reaches temperatures as high as several tens of MeV in young proto-neutron stars born in core-collapse supernovae (CCSN) and in the remnant of a binary neutron star merger (see, e.g., Refs.~\cite{Prakash:1996xs,Oertel:2016bki,FiorellaBurgio:2018dga}). Due to the non-perturbative nature of strong interaction, describing ultra dense and hot matter in their interior remains highly uncertain and model dependent. Confronting model predictions to the large variety of astrophysical observations of CCSN, NS and binary mergers clearly broadens our knowledge on dense matter physics beyond the regimes that can be reached in laboratory experiments. Thereby most observations, such as pulsar masses
or NS tidal deformations from gravitational waves are mainly sensitive to the equation of state (EoS). On the contrary, the signatures from neutrino emission or neutron star's thermal evolution can also help to understand the  various cooling processes, which are very sensitive to matter composition and weak interaction rates.

Let us recall that the role of neutrinos and their interaction with matter in the physics of neutron stars is paramount. First, the birth of neutron stars in CCSN are extremely bright events, with several observations reported well before the modern era of telescopes~\cite{Stephenson2002}. Neutrinos play a crucial role in its dynamics, in particular for the revival of the stalled shock by neutrino heating and a successful explosion within the neutrino-driven mechanism~\cite{Janka2012}. Twenty four neutrinos were detected from the source SN1987a (see Ref.~\cite{Burrows1987} and references therein). For a galactic event with current neutrino detectors, a huge number of detected neutrinos is expected to provide information on the explosion mechanism and on neutrino properties~\cite{Mirizzi:2015eza}. The remnant of this event is a proto-neutron star, which exhibits high temperatures and rapidly cools down through neutrino emission \cite{Prakash:1996xs,Roberts:2012zza,Pascal2022} to become a neutron star. After a few minutes, the star's temperature reaches $\approx 1$\;MeV and the description of neutron star matter can be treated in the zero-temperature limit. This mature neutron star continues to cool down slowly. Neutrino emission is thereby the dominant process for $\approx 10^6$\;years. Last but not least, in a binary neutron star merger remnant the thermodynamic conditions are rather similar to a proto-neutron star with matter being slightly more neutron rich \cite{Endrizzi2020}. Neutrinos are less important for the dynamics of a merger and the evolution of the merger remnant, even if there is energy and momentum exchange with matter \cite{Foucart2023}. They, however, determine the neutron to proton ratio in the ejecta and thus the conditions for the associated heavy element nucleosynthesis, see, e.g., Refs. \cite{Wanajo:2014wha,Martin:2017dhc,Radice:2018pdn,Kullmann:2021gvo,Fujibayashi:2022ftg}. 

There exists a large number of different processes that occur in dense matter contributing to neutrino emission, see Refs.~\cite{Bruenn:1985en,Buras:2005rp,Burrows:2004vq,Pons:1998mm,Yakovlev2001,Page2009,Kiuchi2012,Schmitt2018,Fujibayashi2020, Foucart2023} for discussions of the processes at play within the astrophysical scenarios discussed above. In the present work, we will concentrate on charged-current reactions on nucleons. A recurring process is the so-called direct Urca (DUrca) reaction: a nucleon $N_1$ is converted into a nucleon $N_2$ via the charged-current weak interaction, involving a charged lepton denoted $l^{\pm}$ and an (anti-)neutrino denoted $\bar{\nu_l}/\nu_l$, i.e.,
\begin{align}
    & N_1 \leftrightarrow N_2 + l^{-} + \bar{\nu}_{l} \; , \\
    & N_2 + l^{-} \leftrightarrow N_1 + \nu_{l} \; , \\
    & N_1 + l^{+} \leftrightarrow N_2 + \bar{\nu}_{l} \; , \\
    & N_2 \leftrightarrow N_1 + l^{+} + \nu_{l} \; .
\end{align}
The rates for these reactions have been studied extensively in the literature with different levels of approximation; both in the zero-temperature limit, see, e.g., Refs.~\cite{Friman1979, Lattimer1991,Blaschke1995, Yakovlev1995, Yakovlev2001,Schmitt2018}, and for finite-temperature matter, see, e.g., Refs.~\cite{Bruenn:1985en,Reddy1998,Fu2008,Roberts2017,Alford2018,Fischer2020,Oertel2020,Shin:2023sei}. Due to the kinematics of the reaction, DUrca processes are exponentially suppressed with the inverse of the temperature for certain thermodynamic conditions and (anti-)neutrino energies. In the zero-temperature limit and in $\beta-$equilibrium, i.e., for cold neutron stars, this leads to the well-known DUrca threshold as function of  density. It is the matter's proton fraction that determines whether the very efficient cooling via DUrca is active or not. At finite temperature the threshold is broadened by thermal effects. 

The so-called modified Urca (MUrca)~\cite{Chiu1964} reactions entail a spectator nucleon $N$ interacting \textit{via} strong interaction with the baryons involved in the charged-current weak process, 
\begin{align}
    & N+ N_1 \leftrightarrow N + N_2 + l^{-} + \bar{\nu}_{l} \; , \label{eq:MUrcareactions1}\\
    & N + N_2 + l^{-} \leftrightarrow N + N_1 + \nu_{l} \; , \label{eq:MUrcareactions2}\\
    & N + N_1 + l^{+} \leftrightarrow N + N_2 + \bar{\nu}_{l} \; , \label{eq:MUrcareactions3}\\
    & N + N_2 \leftrightarrow N + N_1 + l^{+} + \nu_{l} \; . \label{eq:MUrcareactions4}    
\end{align}
The neutron and proton branches of this process refer to a neutron or proton as a spectator nucleon, respectively. 

The pioneering study on determining the rate for MUrca-type reactions is that of \textcite{Friman1979}, in which the neutron branch of the reaction is considered for cold matter in $\beta$-equilibrium. Analytic expressions for the rate have been derived under some assumptions: (i) the strong interaction is treated within one-pion exchange (OPE) approximation supplemented with a short-range interaction, (ii) the electron momentum is neglected, and (iii) all involved particles are approximated to be on their respective Fermi surface. Expressions for the proton branch following the same assumptions can be found in Ref.~\cite{Yakovlev1995}. In \textcite{Yakovlev2001} the matrix element is given in the OPE approximation for nonzero electron momenta, keeping the approximations for the phase space in order to still obtain analytic expressions for the rates. 

Since then, much effort has been devoted to investigating different effects in the dense and cold neutron star matter which modify the MUrca rates, see Ref.~\cite{Schmitt2018} for a review. In particular, in Ref.~\cite{Shternin2018} the nucleon-nucleon interaction is treated in the framework of the non-relativistic Bruckner–Hartree–Fock theory and \textit{in medium} effects are included in the energy of nucleons for cold and $\beta$-equilibrated matter. The authors show that the common approach for the propagator of intermediate nucleons approximated to be the electron energy $E_e$ misses an enhancement of the MUrca rate close to the DUrca threshold. 

In contrast to the neutral current bremsstrahlung process, which is known to influence neutrino spectra in CCSN and has thus received some attention, see, e.g., Refs.~\cite{Sigl:1997ga,Hannestad1998,Sedrakian2000,Hanhart:2000ae,Guo:2019cvs}, there are only a few studies on MUrca reactions at finite temperature. In Refs.~\cite{Alford2018, Alford:2021ogv}, the authors include finite temperature effects in an effort to accurately compute Urca neutrino emission processes for the merger of neutron stars and the conditions for achieving $\beta$-equilibrium, but keep the Fermi surface approximation to compute MUrca rates. A purely phenomenological approach for the MUrca rates is applied in Ref.~\cite{Pascal2022}. It is based on the idea that the excitation of two-particles states required to describe reactions in Eqs.~\eqref{eq:MUrcareactions1}-\eqref{eq:MUrcareactions4}, leads to a collisional broadening of the quasi-particles and can thus be simply incorporated as a finite width within DUrca-type reactions~\cite{Roberts:2012um}. 

Because of vector current conservation, the vector current contribution to the neutral current rate vanishes in the limit of zero-momentum transfer, and \textcite{Friman1979} have shown that the vector contribution vanishes within their approximation for the charged-current MUrca reactions too. Therefore it is generally assumed that the axial current contribution dominates and the vector one is neglected.

The aim of this paper is to compute the rates of the MUrca processes for finite temperature neutron star matter composed of neutrons, protons, electrons and positrons ($npe$ matter) lifting most of the above mentioned approximations. We will discuss their respective influence on the rates under different thermodynamic conditions. 

The paper is organized as follows. Section~\ref{sec:Methods} focuses on the analytical derivation of MUrca neutrino emission. First, the formulation to compute the neutrino and anti-neutrino emissivity and mean free path is presented. Then the derivation for the hadronic part of the process is detailed. Common approximations taken to compute the MUrca neutrino rates are also discussed in this section in more detail. In section~\ref{sec:Results}, the results of our calculations are presented. The different components of the hadronic part of the process are discussed and results including the leptons are presented. 

\section{Methods}\label{sec:Methods}

Since MUrca processes involve multiple strong interactions between nucleons, a consistent calculation of the rates with a complete evaluation of in medium effects would be extremely involved. Here, we will evaluate the rates for matter at finite temperature using thermal quantum field theory. We limit ourselves to contributions on the two-loop level, i.e. to contributions related to two-particle two-hole excitations, and do not consider multiple scattering effects. Nucleons will be treated on the mean-field level, such that in medium effects are considered via effective masses and chemical potentials. The matrix element will be derived within the one-pion exchange approximation. We have chosen the latter for better comparison with the existing literature although, as already discussed in \textcite{Friman1979}, a more complete description of the interaction is expected to influence the final rates.  This approach allows us not only to consistently include mean-field effects, is easily generalizable to other prescriptions than OPE for the nucleonic interaction, but in particular, allows us, upon performing the respective approximations, to recover previously established  results~\cite{Friman1979,Shternin2018,Alford2018,Alford:2021ogv}. In Section~\ref{sec:Results/Divergence} we will discuss in more detail its limitations.  

For our numerical computations, we will use the nucleonic equation of state SRO(APR) \cite{Schneider2019} which treats non relativistic particles: the nucleon effective chemical potentials and effective masses of baryons as a function of the temperature, density and lepton fraction are provided in the table of the \href{https://compose.obspm.fr/eos/149}{CompOSE} online database \cite{aprCompose,Typel:2013rza,CompOSECoreTeam:2022ddl}, and both quantities will be used in our approach to account for \textit{in medium} effects.

\subsection{Neutrino emissivity and mean-free path of modified Urca reactions}\label{sec:Methods/neutrinoEmission}

The neutrino emissivity and mean free path can be established from the time derivative of the neutrino and anti-neutrino distribution functions denoted $\mathcal{F}_{\nu}$ and $\mathcal{F}_{\bar{\nu}}$ respectively, according to 
\begin{align}
    \frac{\partial}{\partial_t} \mathcal{F}_{\nu} &= j(E_{\nu}) \big( 1 - \mathcal{F}_{\nu}\big) - \frac{1}{\lambda(E_{\nu})} \mathcal{F}_{\nu}\;, \\
    \frac{\partial}{\partial_t} \mathcal{F}_{\bar{\nu}} &= \bar{j}(E_{\nu}) \big( 1 - \mathcal{F}_{\bar{\nu}}\big) - \frac{1}{\bar{\lambda}(E_{\nu})} \mathcal{F}_{\bar{\nu}} \;,
\end{align}
with $E_{\nu}$ the energy of massless (anti-)neutrinos. The neutrino and anti-neutrino emissivity are denoted $j$ and $\bar{j}$, respectively, and the neutrino and anti-neutrino mean-free paths are denoted $\lambda$ and $\bar{\lambda}$, respectively. Details of the derivation can be found, e.g., in Refs.  \cite{Sedrakian:1999jh,Sedrakian:2006mq,Schmitt2006} or Section II.~A of Ref.~\cite{Oertel2020}. In many applications the absorption opacity corrected for stimulated absorption is introduced,
see, e.g., Ref.~\cite{Burrows:2004vq},
\begin{align}
  \kappa_a^*(E_\nu) &= \frac{1}{1 - \mathcal{F}^{eq}(E_\nu-\mu_\nu)} \frac{1}{\lambda(E_\nu)} = j(E_\nu) + \frac{1}{\lambda(E_\nu)}~, \nonumber \\
  \bar{\kappa_a^*}(E_\nu) &= \frac{1}{1 - \mathcal{F}^{eq}(E_\nu+\mu_\nu)} \frac{1}{\bar{\lambda}(E_\nu)} = \bar{\jmath}(E_\nu) + \frac{1}{\bar{\lambda}(E_\nu)}~,\nonumber \\
\end{align}
where $\mathcal{F}^{eq}(E_\nu \pm \mu_\nu)$ is the (anti-)neutrino distribution function at equilibrium (Fermi-Dirac), with $\mu_\nu$ the neutrino chemical potential. Speaking about opacities below, we will always refer to $\kappa_a^*$ and $\bar{\kappa}_a^*$ for neutrinos or
anti-neutrinos, respectively. In our derivation, neutrinos and anti-neutrinos are treated independently: no interaction between neutrinos and anti-neutrinos, nor neutrino oscillations are taken into account.

In this paper, we focus on the MUrca processes involving neutrons $n$, protons $p$, electrons $e^-$, and positrons $e^+$; $npe$ matter involves the following eight reactions:
 \begin{align}
    &N + n \leftrightarrow N + p + e^- + \bar{\nu}\;, \label{eq:MUrca1} \\
    &N + p + e^- \leftrightarrow N + n + \nu \;, \label{eq:MUrca2} \\
    &N + n + e^+ \leftrightarrow  N + p + \bar{\nu} \;, \label{eq:MUrca3} \\
    &N + p \leftrightarrow  N + n + e^+ + \nu \;,\label{eq:MUrca4} 
\end{align}
with $N$ the spectator nucleon. Modified neutron decay and its inverse are presented in Eq.~\eqref{eq:MUrca1}, modified electron capture and its inverse are presented in Eq.~\eqref{eq:MUrca2}, modified positron capture and its inverse are presented in Eq.~\eqref{eq:MUrca3}, and modified proton decay and its inverse are presented in Eq.~\eqref{eq:MUrca4}. In the following, each pair of reactions is designated by an index referring to the leptons involved: $[e^- \bar{\nu}]$, $[e^- \nu]$, $[e^+ \bar{\nu}]$ and $[e^+ \nu]$. Both the neutrino and the anti-neutrino opacity are contributed to by processes involving electrons and positrons.

Explicitly, the emissivities for the different reactions (see, e.g., Ref.~\cite{Oertel2020}) are given by 
\begin{align}
    j^{[e^- \nu]} & = - \frac{G_F^2 V_{ud}^2}{8} \int \frac{{\rm d}^3 \vec{p}}{(2 \pi)^3}  \frac{L_{\alpha \beta} (Q^{[e^{-}\nu]}) \Im \Pi^{\alpha \beta} (Q^{[e^{-}\nu]}) }{E_e E_{\nu}}\nonumber \\ 
    & \hspace{1.5cm} \times \mathcal{F}_{e^-} \big(1 + \mathcal{F}_B(Q_0^{[e^-\nu]}) \big)  \;,\label{eq:j1} \\
    j^{[e^+ \nu]} & = \frac{G_F^2 V_{ud}^2}{8} \int \frac{{\rm d}^3 \vec{p}}{(2 \pi)^3}  \frac{L_{\alpha \beta} (Q^{[e^{+}\nu]}) \Im \Pi^{\alpha \beta} (Q^{[e^{+}\nu]}) }{E_e E_{\nu}}  \nonumber \\ 
    & \hspace{1.5cm} \times \big( 1- \mathcal{F}_{e^+} \big) \big(1 + \mathcal{F}_B(Q_0^{[e^+\nu]}) \big) \;, \label{eq:j2}  \\
    \bar{j}^{[e^- \bar{\nu}]} & = \frac{G_F^2 V_{ud}^2}{8} \int \frac{{\rm d}^3 \vec{p}}{(2 \pi)^3}  \frac{L_{\alpha \beta} (Q^{[e^{-}\bar{\nu}]}) \Im \Pi^{\alpha \beta} (Q^{[e^{-}\bar{\nu}]})}{E_e E_{\nu}} \nonumber \\ 
    & \hspace{1.5cm} \times \big( 1- \mathcal{F}_{e^-} \big) \mathcal{F}_B(Q_0^{[e^-\bar{\nu}]}) \;, \label{eq:j3} \\
    \bar{j}^{[e^+ \bar{\nu}]} & = -\frac{G_F^2 V_{ud}^2}{8} \int \frac{{\rm d}^3 \vec{p}}{(2 \pi)^3}  \frac{L_{\alpha \beta} (Q^{[e^{+}\bar{\nu}]}) \Im \Pi^{\alpha \beta} (Q^{[e^{+}\bar{\nu}]})}{E_e E_{\nu}} \nonumber \\ 
    & \hspace{1.5cm} \times \mathcal{F}_{e^+} \mathcal{F}_B(Q_0^{[e^+ \bar{\nu}]})   \;, \label{eq:j4} 
\end{align}
with $\vec{p}$ and $E_e$ the charged lepton (three-)momentum and energy, $\mathcal{F}_B$ the Bose-Einstein distribution function, $L$ the lepton tensor and $\Pi$ the retarded hadronic polarization function, $G_F$ the Fermi coupling constant and $V_{ud}$ the up/down component of the electroweak mixing matrix. Analogous expressions for the inverse mean free path can be obtained from detailed balance relations. 

The four-momentum denoted ${Q=(Q_0, \vec{Q})}$ corresponds to energy and momentum exchanged in the reaction; its dependence on the individual particle's momenta depends on the reaction at hand. We denote $|\vec{Q}|$ the momentum norm and the minimum or maximum possible value for transferred energy is denoted $Q_{0;\rm ext}$ such that
\begin{align}\label{eqQq0}
     & \begin{cases}
        |\vec{Q}|^2 = (\vec{p} + s_1 s_2 \, \vec{q} )^2\;, \\
        Q_0 = s_1 E_e + s_2 E_{\nu} - \mu_e + \mu_{\nu} \;, \\
        Q_{0; \rm ext} = s_1m_e + s_2E_{\nu} - \mu_e + \mu_{\nu} \;, 
    \end{cases} 
\end{align}
with $\mu_e$ the electron chemical potential and $\vec{q}$ the (anti-)~neutrino momentum with $|\vec{q}| = E_\nu$. The signs $s_1,s_2$ for the different reactions are given by 
\begin{align*}
    (s_1,s_2) =& \begin{cases}
        (+1,+1) \; \text{for Eq.~\eqref{eq:MUrca1}} \;, \\
        (+1,-1) \; \text{for Eq.~\eqref{eq:MUrca2}} \;, \\
        (-1,+1) \; \text{for Eq.~\eqref{eq:MUrca3}} \;, \\
        (-1,-1) \; \text{for Eq.~\eqref{eq:MUrca4}} \;.
    \end{cases}  
\end{align*}

The lepton tensor that appears in the expression of the neutrino emissivities in Eqs.~\eqref{eq:j1}-\eqref{eq:j4} reads
\begin{equation*}
    L_{\alpha \beta} = 8 \big( p_{\alpha} q_{\beta} + q_{\alpha} p_{\beta} - p \cdot q \; g_{\alpha \beta} - {\rm i} \epsilon_{\rho \lambda \alpha \beta } p^{\rho} q^{\lambda} \big) \;, \label{eq:componentsLeptonTensor}
\end{equation*}
with $g$ the Minkowski metric in $(+,-,-,-)$ signature, $\epsilon$ the Levi-Civita symbol, $p$ the charged lepton four-momentum, and $q$ the (anti-)neutrino four-momentum.

\subsection{Hadronic polarization function} \label{sec:Methods/hadronFunction}

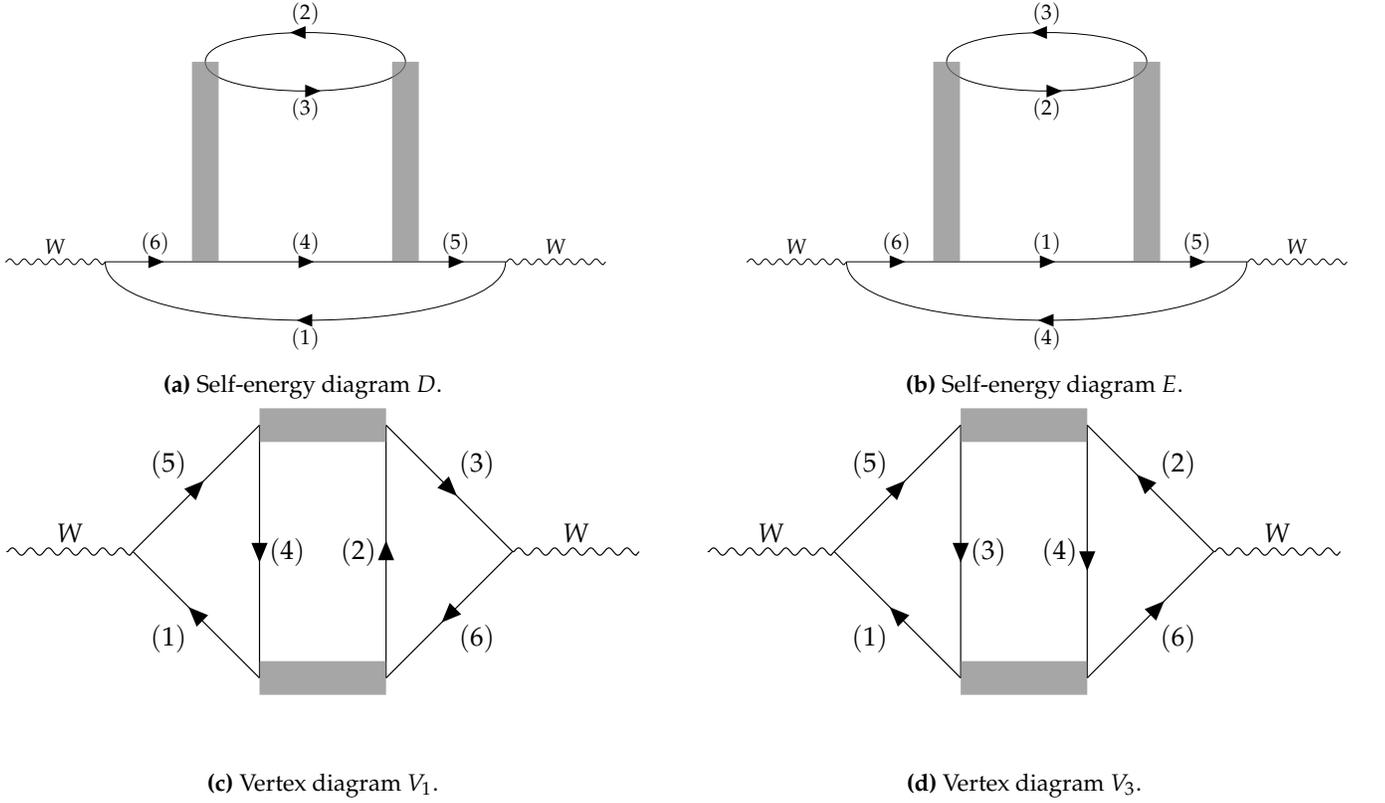
\begin{figure*}
\begin{subfigure}{.45\textwidth}
  \resizebox{\hsize}{!}{
  \begin{tikzpicture}
  \begin{feynman}
  \vertex (a); 
  \vertex [right =of a] (b); 
  \vertex [right =of b] (c);
  \vertex [right =of c] (d);
  \vertex [right =of d] (e);
  \vertex [above =of b] (f); 
  \vertex [above =of f] (g);
  \vertex [above =of d] (h); 
  \vertex [above =of h] (i);
  \vertex [left = of a] (k); 
  \vertex [right = of e] (l);
  \diagram* {
        (k) -- [photon, edge label = \(W\)] (a),
        (a) -- [fermion,  edge label=\((6)\)] (b) -- [fermion, edge label=\((4)\)] (d) -- [fermion, edge label=\((5)\)] (e), 
        (e) -- [fermion, half left, looseness=0.5,edge label=\((1)\)] (a),
        (i) -- [anti fermion, half left, looseness=0.5, edge label=\((3)\)] (g),
        (g) -- [anti fermion, half left, looseness=0.5, edge label=\((2)\)] (i), 
        (e) -- [photon, edge label = \(W\)] (l),
    };
    \end{feynman}
    \draw [gray, line width=4mm, opacity=.7] (b) -- node[below=5mm,font=\huge] {} (g);
    \draw [gray, line width=4mm, opacity=.7] (d) -- node[below=5mm,font=\huge] {} (i);
\end{tikzpicture}}
  \caption{Self-energy diagram $D$.}
\end{subfigure}
\hfill
\begin{subfigure}{.45\textwidth}
  \resizebox{\hsize}{!}{
  \begin{tikzpicture}
  \begin{feynman}
  \vertex (a); 
  \vertex [right =of a] (b); 
  \vertex [right =of b] (c);
  \vertex [right =of c] (d);
  \vertex [right =of d] (e);
  \vertex [above =of b] (f); 
  \vertex [above =of f] (g);
  \vertex [above =of d] (h); 
  \vertex [above =of h] (i);
  \vertex [left = of a] (k); 
  \vertex [right = of e] (l);
  
  \diagram* {
        (k) -- [photon, edge label = \(W\)] (a),
        (a) -- [ fermion,  edge label=\((6)\)] (b) -- [fermion, edge label=\((1)\)] (d) -- [fermion, edge label=\((5)\)] (e), 
        (e) -- [fermion, half left, looseness=0.5,edge label=\((4)\)] (a),
        (i) -- [anti fermion, half left, looseness=0.5, edge label=\((2)\)] (g),
        (g) -- [anti fermion, half left, looseness=0.5, edge label=\((3)\)] (i), 
        (e) -- [photon, edge label = \(W\)] (l),
    };
    \end{feynman}
    \draw [gray, line width=4mm, opacity=.7] (b) -- node[below=5mm,font=\huge] {} (g);
    \draw [gray, line width=4mm, opacity=.7] (d) -- node[below=5mm,font=\huge] {} (i);
\end{tikzpicture}}
  \caption{Self-energy diagram $E$.}
\end{subfigure}
\newline 
\begin{subfigure}{.48\textwidth}
  \resizebox{\hsize}{!}{
  \begin{tikzpicture}
  \begin{feynman}
  \vertex (a); 
  \vertex [right =of a] (b); 
  \vertex [right =of b] (c);
  \vertex [right =of c] (d);
  \vertex [right =of d] (e);
  \vertex [right =of e] (f); 
  \vertex [below =of c] (g);
  \vertex [below =of d] (h); 
  \vertex [above =of c] (i);
  \vertex [above =of d] (j); 
  \diagram* {
        (a) -- [photon, edge label = \(W\)] (b),
        (b) -- [fermion,  edge label=\((5)\)] (i), 
        (i) -- [fermion,  edge label=\((4)\)] (g), 
        (g) -- [fermion,  edge label=\((1)\)] (b), 
        (j) -- [fermion,  edge label=\((3)\)] (e), 
        (e) -- [fermion,  edge label=\((6)\)] (h), 
        (h) -- [fermion, edge label = \((2)\)] (j),
        (e) -- [photon, edge label = \(W\)] (f),
    };
    \end{feynman}
    \draw [gray, line width=4mm, opacity=.7] (i) -- node[below=5mm,font=\huge] {} (j);
    \draw [gray, line width=4mm, opacity=.7] (g) -- node[below=5mm,font=\huge] {} (h);
\end{tikzpicture}
  }
  \caption{Vertex diagram $V_1$.}
\end{subfigure}
\hfill 
\begin{subfigure}{.48\textwidth}
  \resizebox{\hsize}{!}{
  \begin{tikzpicture}
  \begin{feynman}
  \vertex (a); 
  \vertex [right =of a] (b); 
  \vertex [right =of b] (c);
  \vertex [right =of c] (d);
  \vertex [right =of d] (e);
  \vertex [right =of e] (f); 
  \vertex [below =of c] (g);
  \vertex [below =of d] (h); 
  \vertex [above =of c] (i);
  \vertex [above =of d] (j); 
  \diagram* {
        (a) -- [photon, edge label = \(W\)] (b),
        (b) -- [fermion,  edge label=\((5)\)] (i), 
        (i) -- [fermion,  edge label=\((3)\)] (g), 
        (g) -- [fermion,  edge label=\((1)\)] (b), 
        (j) -- [anti fermion,  edge label=\((2)\)] (e), 
        (e) -- [anti fermion,  edge label=\((6)\)] (h), 
        (h) -- [anti fermion, edge label = \((4)\)] (j),
        (e) -- [photon, edge label = \(W\)] (f),
    };
    \end{feynman}
    \draw [gray, line width=4mm, opacity=.7] (i) -- node[below=5mm,font=\huge] {} (j);
    \draw [gray, line width=4mm, opacity=.7] (g) -- node[below=5mm,font=\huge] {} (h);
\end{tikzpicture}
  }
  \caption{Vertex diagram $V_3$.}
\end{subfigure}
\caption{Feynman representation of MUrca diagrams similar to what is represented in  Fig.~3 of Ref.~\cite{Bacca2012}. Plain lines represent nucleons numbered from $(1)$ to $(6)$, and gray boxes represent the interaction approximated here by one-pion exchange. }
\label{fig:murcadiagrams}
\end{figure*}

The hadronic polarization function can be understood as a weak boson (in the case of the charged current a $W$ boson) self-energy. The lowest order contribution thereby corresponds to a DUrca reaction~\cite{Schmitt2006} and for MUrca reactions two types of contributions must be taken into account: corrections due to a nucleonic self-energy insertion and corrections to the vertex relating the weak boson and the nucleons. We can distinguish two self-energy diagrams later on referred to as diagrams $D$ and $E$, and three vertex-correction diagrams later on referred to as diagrams $V_1$, $V_2$, and $V_3$, see Fig.~\ref{fig:murcadiagrams}.  Vertex $V_2$ is not shown since it is obtained from $V_1$ simply by exchange of particles labeled (3) and (4), respectively, and its contribution is equal to that of $V_1$. The retarded polarization function can thereby be decomposed in a very general way into a vector component denoted $\Pi_V$, an axial transverse component denoted $\Pi_T$, an axial longitudinal component denoted $\Pi_L$, and a mixed component denoted $\Pi_{AV}$ as
\begin{align}
  \Pi^{00}(Q) &=  c_V^2 \Pi_V(Q) \;, \nonumber \\
  \Pi^{ij}(Q) &= c_A^2 \left(\delta^{ij} - \frac{Q^i Q^j}{\vec{Q}^2 } \right)\Pi_T(Q) + c_A^2 \frac{Q^i Q^j}{\vec{Q}^2} \Pi_L(Q) \;, \nonumber \\
  \Pi^{i0}(Q) &=  {\rm i} c_A c_V \Pi_{AV}(Q) \frac{Q^i}{|\vec{Q}|} \;, \nonumber  \\
  \Pi^{0j}(Q) &=  {\rm i} c_A c_V \Pi_{AV}(Q) \frac{Q^j}{|\vec{Q}|} \;, \label{eq:pi}
\end{align}
with $c_A$ and $c_V$ the axial and vector nucleon form factors respectively, which we assume to be constants. For the interactions in play, the mixed contribution vanishes and we will not discuss it any further. Upon taking the imaginary part in order to evaluate the opacities, there are cuts, which correspond to corrections related to the DUrca processes and cuts corresponding to genuine MUrca reactions with two incoming and two outgoing nucleons as well as two intermediate (virtual) nucleons. In the following we will only consider the latter cuts.

As mentioned above, we treat the strong interaction between nucleons indicated as gray boxes in Fig.~\ref{fig:murcadiagrams} within the one-pion exchange approximation; note that this approach might overestimate the matrix element by a factor of a few as found for cold $\beta$-equilibrated neutron star matter, see Ref.~\cite{Schmitt2018} and references therein. The corresponding matrix $T$, is given as
\begin{equation}
    T_{ab;cd}^{ij} (p_1,p_3;p_2,p_4) = -\widebar{T}_{ab}(p_1,p_3) \tau^i \times \widebar{T}_{cd}(p_2,p_4) \tau^j \;,
\end{equation}
with $\tau$ the Pauli matrices associated to isospin, $p_k$ designating the nucleonic four-momenta, and the spin indices denoted $\{a,\dots,d \}$. The nucleons are treated here as non relativistic particles, which should be justified since their masses are much higher than typical energies. As a remark of caution, let us mention that relativistic corrections were shown to become important already at moderate densities the DUrca process~\cite{Leinson:2001ei,Leinson:2002bw,Alford2022}. We, however, keep the non-relativistic approximation for simplicity. The quantity $\widebar{T}$ is given by 
\begin{equation}
    \widebar{T}_{ab}(\vec{p}_k,\vec{p}_l) =  \frac{f_{\pi NN}}{m_{\pi}}\left[ \frac{\vec{\sigma} \cdot (\vec{p}_l-\vec{p}_l) }{\sqrt{(\vec{p}_l-\vec{p}_k)^2+m_{\pi}^2}} \right]_{ab}  \;.
\end{equation}
with $\sigma$ the Pauli matrix associated to spin, $f_{\pi NN}$ the pion-nucleon-nucleon coupling constant, and $m_{\pi}$ the pion mass; for numerical applications, we will take ${f_{\pi NN} = 1.01}$ and ${m_\pi = 140}$\;MeV~\cite{Rapp:1997ei}. Considering all possible direct and exchange contributions for the couplings of the nucleons, there are four different contributions to each diagram. 

We then write the nucleon propagators  as
\begin{equation}
    {\mathcal{S}}^{ab}(k) = \frac{1}{2} \sum_{x=n,p} S^{ab}_x(k) \big( \mathbb{1}_2 + t_x \tau^3 \big) \; ,
\end{equation}
with $t_x$ the third component of isospin with the convention that for neutrons $t_n=+1$ and for protons $t_p=-1$, $k$ is the four-momentum of the nucleon, $\mathbb{1}_2$ is the two-dimensional identity matrix. In the mean-field approximation, the nucleon propagator for each isospin assumes the form of a free one, 
\begin{equation}
     S_x^{ab}(k) = \frac{\delta^{ab} }{k_{0;x} - \mathscr{E}_{\vec{k};x} } \;, \label{eq:propNucleons}
\end{equation}
with the particle energies denoted as $\mathscr{E}_{\vec{k};x}$ given by
\begin{equation}
    \mathscr{E}_{\vec{k};x} = E_x(k) - \mu^*_x = \frac{\vec{k}^2}{2 m^*_x} + m^*_x - \mu^*_x \;, 
\label{eq:meanfieldpropagator}
\end{equation}
containing effective masses $m^*_x$ and chemical potentials $\mu^*_x$, which describe the in-medium effects on the nucleon propagators due to interactions on the mean-field level. 

Finally, the last element that appears in the expression of the hadronic polarization function is the vertex between the weak boson and nucleon-nucleon coupling denoted $\Gamma$ and given by
\begin{equation}
    \Gamma_{\pm;ab} = \big[(c_V + c_A \vec{\sigma} ) \tau_{\pm} \big]_{ab}  \; , 
\end{equation}
with $\tau_{\pm} =  1/2 (\tau^1\pm {\rm i}\tau^2)$\;.

In practice, we will calculate first the temperature polarization function in Matsubara formalism and then use analytical continuation to obtain the retarded one, see Appendix~\ref{Appendix:matsuFunction} for details. It can then be written in the following form: 
\begin{widetext}
    \begin{align}
        \Im \Pi^{\alpha \beta} (Q) &= \bigg( \prod_{j = 1}^4 \int \frac{{\rm d}^3 \vec{p}_j}{(2\pi)^4} \bigg)  \sum_{X=D,E,V_1,V_2, V_3}  \sum_{I(X)} \Im \mathcal{M}_X^{I(X)} (Q,\vec{p}_1, \vec{p}_2, \vec{p}_3, \vec{p}_4)\nonumber \\ 
        & \hspace{1cm} \times \mathcal{X}_{X}^{\alpha \beta}(Q,\vec{p}_1, \vec{p}_2, \vec{p}_3, \vec{p}_4)  \delta( \vec{Q} + \vec{p}_{1} + \vec{p}_2 - \vec{p}_3 - \vec{p}_4 ) \;, \label{eq:integral}
    \end{align}
\end{widetext}
where the explicit form of each term is obtained by performing the traces in spin and isospin space for each diagram, and then the sum over Matsubara frequencies. 
The function $\mathcal{M}$ thereby contains different combinations of the nucleonic distribution functions for two incoming and two outgoing nucleons, see Appendix~\ref{Appendix:matsuFunction} for explicit expressions. It is also affected by the various isospin combinations $I(X)$ associated to each diagram ${\{X = D,E,V_1, V_2,V_3\}}$, see Appendix~\ref{Appendix:isospin} for details. The tensor $\mathcal{X}$ comprises the results from the trace in spin space for the different diagrams including the four different direct and exchange contributions $\{a,\dots,d\}$, it is detailed in Appendix~\ref{Appendix:spin}. This complete expression is integrated numerically using a Monte-Carlo method with importance sampling, for which details are discussed in Appendix~\ref{Appendix:MC}. 

\subsection{Common approximations}\label{sec:Methods/approximations}
As discussed in the Introduction, rates for MUrca processes have been previously obtained in the literature, but only applying several approximations to the full expression in Eq.~\eqref{eq:integral} in order to simplify the calculations. Let us start with the two most commonly used ones.

\subsubsection{Propagator of intermediary particles}\label{sec:Methods/approximations/propagator}

The functions $\mathcal{M}$ presented in the Appendix~\ref{Appendix:matsuFunction} contain a function $\mathscr{S}_X$, which has the typical structure
\begin{align}
    \mathscr{S}_X &= 
    \frac{1}{\big( {\mathscr{E}}_{\vec{p}_{k};w} -{\mathscr{E}}_{\vec{p}_l;x} - Q_0 \big)
    \big( {\mathscr{E}}_{\vec{p}_{m};y} -{\mathscr{E}}_{\vec{p}_n;z} - Q_0 \big)}  \;,  \label{eq:denominator}
\end{align}
from the propagators of intermediate nucleons involved in the DUrca cuts of the respective diagrams. It is common in the literature to use a simple approximation for $\mathscr{S}$, the same for all diagrams: 
\begin{itemize}
    \item $\mathscr{S} = 1/E_e^2$, see, e.g., Refs.~\cite{Friman1979,Yakovlev1995,Yakovlev2001, Alford2018}\;,
    \item $\mathscr{S} = 1/Q_0^2$, see, e.g., Ref.~\cite{Bacca2012} \;.
\end{itemize}
The former is justified by the fact that upon neglecting mean-field effects and the neutrino energy, the difference of the two nucleon energies in Eq.~\eqref{eq:denominator} reduces to ${\approx \mu_p - \mu_n}$ and the entire denominator thus to $\approx E_e^2$, the energy of the charged lepton squared. In Ref.~\cite{Bacca2012}, the authors perform their calculations in pure neutron matter and neglect the momentum exchange, i.e. $\vec{Q} = 0$, such that naturally the denominator in Eq.~\eqref{eq:denominator} reduces to $Q_0$. We will discuss the validity of these approximations in Section~\ref{sec:Results}.

\subsubsection{Fermi surface and $\beta$-equilibrium approximations}\label{sec:Methods/approximations/Fermi+beta}

In the seminal work of \textcite{Friman1979} and many subsequent studies, nucleons and charged leptons are assumed to reside on their respective Fermi surface which tremendously simplifies the phase space integrations in Eq.~\eqref{eq:integral} and Eqs.~\eqref{eq:j1}-\eqref{eq:j4}. The Fermi surface is well defined in the zero-temperature limit and this approximation is then perfectly valid. Thermal broadening of the Fermi-Dirac distribution at finite temperature, however, complicates the definition of the Fermi surface and renders the approximation for the phase space integration questionable. However, for the sake of obtaining an analytical expression for MUrca rates, it is still applied in the literature. The distribution function of a nucleon $\mathcal{F}_{F;x}$ of isospin nature $x$ is then still reduced to a Heaviside step function with the Fermi momentum of the particle denoted $p_{F,x}$ defined as, see, e.g., Refs~\cite{Alford2018,Alford:2021ogv}
\begin{equation}
    p_{F,x} = (3 \pi^2 n_x)^{1/3} \;, 
\end{equation}
with $n_x$ the number density of the nucleon. 
The individual number densities are thereby determined from the EoS with given thermodynamic conditions. Assuming matter in $\beta$-equilibrium simplifies the calculations because at a given temperature and density, it determines the charged lepton (electron) fraction; together with charge neutrality, the Fermi momenta of the different particles species are effectively related. Most of the time the $\beta$-equilibrium condition is thereby implemented assuming matter to be transparent to neutrinos, i.e., with vanishing neutrino equilibrium chemical potential, see Ref.~\cite{Alford2018} for corrections in the case of a non-zero $\mu_\nu$.

\subsubsection{Additional approximations}
Many works apply additional approximations on the phase space as well as on the matrix element. Most commonly the matrix element is determined neglecting the momentum exchange $\vec{Q}$. This arises as a combination of two assumptions: first neutrino momentum is considered as much smaller as all other momenta and then electron momentum is neglected for the matrix element. In this case, the vector contribution to the hadronic polarization function vanishes in the OPE approximation as already shown in \textcite{Friman1979}. Together with the argument of vector current conservation valid for the corresponding neutral current bremsstrahlung processes, most of the time only the axial part is considered.
Additionally, for the phase space, very often an averaging is performed over the  orientations of the neutrino momentum. Combining the above assumptions,  analytical expressions of the MUrca rates can be obtained, see \textcite{Yakovlev2001}; in Ref.~\cite{Kaminker2016} details on angular integrations can be found. 

A last point to be mentioned here is that it has been shown for DUrca processes that including mean-field effects via effective masses and chemical potentials has an important effect on the rates, see Refs.~\cite{Roberts:2012um,Martinez-Pinedo:2012eaj}. The reason is that in asymmetric matter, the interaction potential entering the effective chemical potentials is different for neutrons and protons, thus modifying the energy difference between neutron and proton states, and consequently the available energy for a charged-current process. A similar reasoning applies to MUrca reactions in asymmetric matter, but up to now effective masses and chemical potentials have not been largely considered in determining MUrca rates. An exception is the work by \textcite{Alford:2021ogv}, which includes these effects and discusses using relativistic kinematics for the single-particle energies, which can become important at high densities upon the effective masses being very small in many equation of state models.

\section{Results}\label{sec:Results}

\subsection{Intermediary propagator above the DUrca threshold} \label{sec:Results/Divergence}

\begin{figure}
    \centering
    \resizebox{!}{\hsize}{\includegraphics{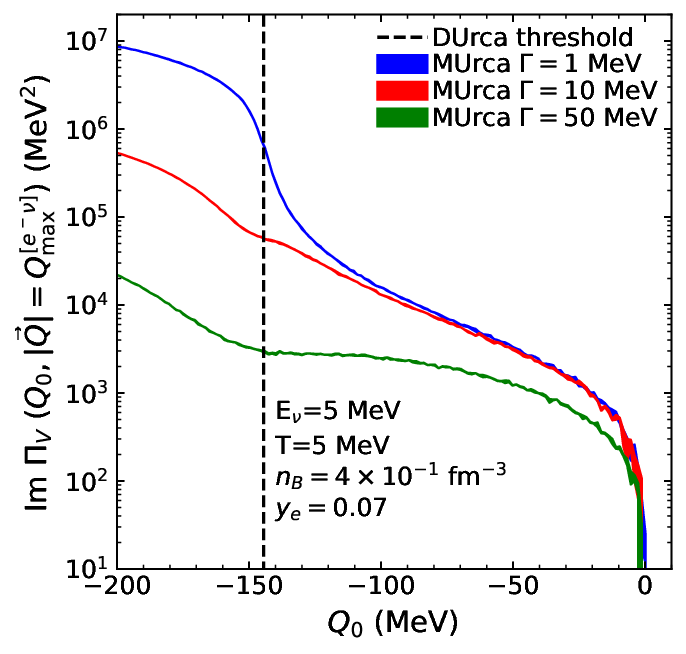}}
    \caption{Vector component of the imaginary part of the hadronic polarization function as a function of energy exchange $Q_0$, for the maximum value of $|\vec{Q}|$ allowed in the case of an electron capture. This quantity is represented for different values of the parameter $\Gamma$ (see text for details). Results are presented for a Monte-Carlo method with $5\times10^5$ points and the numerical error of this approach is represented in the width of the line (see Appending~\ref{Appendix:MC}).}
    \label{fig:gammaTest}
\end{figure}
Before discussing the results for the hadronic polarization function, we must first discuss its value close to the direct Urca threshold in our approach. As already mentioned in the Introduction, in Ref.~\cite{Shternin2018} the authors show that approximating the denominator of the intermediate nucleons propagator to the squared charged lepton energy $E_e^2$ misses an enhancement of the MUrca rate close to the DUrca threshold. This effect can be understood as follows: the structure of the denominators arising from the propagators of the intermediate nucleons, see Eq.~\eqref{eq:denominator}, clearly shows that above the corresponding DUrca threshold, the intermediate particles can become on-shell, leading to poles in the contribution to MUrca processes. Close to the threshold an enhancement is observed as in \textcite{Shternin2018} because the denominator approaches the poles. However, the divergence is an artifact of our treatment considering nucleons in mean-field approximation as intermediate particles. If we had correctly included dressed nucleons with a non-zero lifetime, the divergence would disappear. Performing this operation correctly is a formidable task which we will not attempt to do here. Instead we choose a simple way to circumvent the divergence inspired by the above reasoning: we introduce a parameter $\Gamma$ in the intermediate propagators of all diagrams. It would correspond to a constant imaginary part of a nucleon self-energy in the intermediate particle propagators of the vertex correction diagrams. The function $\mathscr{S}$ in Eq.~\eqref{eq:denominator} then assumes the form 
\begin{align}
     \mathscr{S}_X = \frac{\big( \mathscr{E}_{A} -\mathscr{E}_{B} \pm Q_0 \big)\big( \mathscr{E}_{C} -\mathscr{E}_{D} \pm Q_0 \big)}{\big[ \big( \mathscr{E}_{A} -\mathscr{E}_{B} \pm Q_0 \big)^2 + \Gamma^2 \big] \big[ \big( \mathscr{E}_{C} -\mathscr{E}_{D} \pm Q_0 \big)^2 + \Gamma^2 \big]}  \;, \label{eq:intermediaryProp}
\end{align}
which reduces to Eq.~\eqref{eq:denominator} in the limit $\Gamma \to 0$. A similar strategy has been followed in Ref.~\cite{Hannestad1998} for the neutral current bremsstrahlung process to cure the divergence which appears at $Q_0 \to 0$ in that case. The authors have been able to determine the value of $\Gamma$ from sum rules, which unfortunately do not apply in our case. The transport approach developed for neutral currents in Ref.~\cite{Lykasov:2008yz,Bacca2012}, allows us to include consistently the collisional broadening due to multiple scattering effects within a quasi-particle lifetime for the nucleons. This approach has been implemented in a purely phenomenological way to the charged-current MUrca reactions in Ref.~\cite{Pascal2022} following \textcite{Roberts:2012um}. A consistent generalization of this approach to the charged current could indeed be a promising possibility to avoid the artifact and the divergence, but is beyond the scope of the present paper, see Ref.~\cite{Bartl2016} for a first attempt. Here, awaiting for a consistent way to determine its value, see also Refs.~\cite{Voskresensky:1987hm,Knoll:1995nz}, our results will depend on the value chosen for $\Gamma$. We will choose a value, which allows us to compute results around and beyond the DUrca threshold while not affecting the calculation far from the threshold.

To that end, we have performed computations with different values of $\Gamma$. As an example, in Fig.~\ref{fig:gammaTest}, the vector component of the imaginary part of the hadronic polarization function is displayed for a neutrino energy of 5\;MeV\;; the thermodynamic conditions are $T=50$\;MeV for the temperature, $n_B = 0.4$\;fm$^{-3}$ for the baryon density and $Y_e=0.07$ for the electron fraction.  In Fig.~\ref{fig:gammaTest}, $\Im \Pi_V$ is presented for $\Gamma = 1, 10, 50$\;MeV and a range of energy transfers ${Q_0=[-200:10]}$\;MeV in which kinetics are in favor of electron captures and neutron decays.  The momentum transfer $|\vec{Q}|$ has been chosen at the maximum value ${|\vec{Q}| = Q_{\rm ext}^{e^- \nu} = Q_{\rm max}}$ for an electron capture reaction. The DUrca threshold is represented by a black vertical dashed line at $Q_0=-150$\;MeV. In the range $Q_0=[-200:-150]$\;MeV, close to the DUrca threshold, the three curves differ by orders of magnitude, and the parameter $\Gamma$ plays a role in the results. This is also the case in the range $Q_0=[-150:100]$\;MeV. However, for larger $Q_0$, the curve for $\Gamma = 1$\;MeV and $\Gamma=10$\;MeV join, while the curve for $\Gamma=50$\;MeV remains different. For the remaining results, we will set the parameter to $\Gamma =10$\;MeV.

\subsection{Different components of the hadronic polarization function}

\begin{figure}
    \centering
    \resizebox{!}{\hsize}{\includegraphics{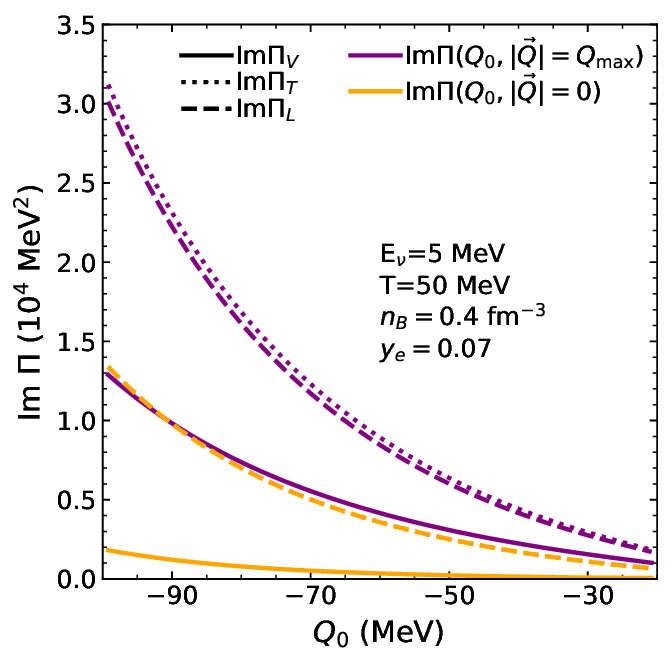}}
    \caption{Vector, axial longitudinal and axial transverse components of the imaginary part of the polarization function as a function of $Q_0$. Results are presented for $|\vec{Q}|= Q_{\rm ext}^{e^- \nu} = Q_{\rm max}$ and for $|\vec{Q}|=0$. They were established with a Monte-Carlo approach with $5\times 10^5$ points; for clarity in this figure, the numerical error is not represented and a fit has been performed for each curve; the numerical error are of the same order as that presented in Fig.~\ref{fig:gammaTest}.}
    \label{fig:axVsvec}
\end{figure}

The hadronic polarization function can be decomposed into a vector component, a transverse and a longitudinal axial component, and a mixed vector-axial component, see Eq.~\eqref{eq:pi}. As mentioned before, most evaluations ignore the vector component and only compute the transverse axial one. Here we want to assess to which extent this assumption is valid. The mixed contribution vanishes and we will not consider it further here. 

The vector component and the two axial components of the imaginary part of the hadronic polarization function, $\Im \Pi$,  are presented in Fig.~\ref{fig:axVsvec}. The axial components are of the same order and dominate the vector component by a factor of a few. In Fig.~\ref{fig:axVsvec}, results for the axial and vector components of the imaginary part of the hadronic polarization function are presented for a momentum transfer $|\vec{Q}|=Q_{\rm max}$ and for $|\vec{Q}|=0$. In the former case it is evident that the vector component cannot be neglected. In the latter case, results for both the axial and vector components are reduced by a factor of $\approx 2.5$ compared to $|\vec{Q}|=Q_{\rm max}$, but the vector component is not strictly vanishing, as would have been expected from the results in \textcite{Friman1979}. It can indeed be shown that the sum of all diagrams for the vector component vanishes in the limit $\vec{Q} \to 0$ if in addition the denominator of the intermediate particles is assumed the same for all diagrams. This is the case in \textcite{Friman1979}, where it has been assumed to be $\approx E_e^2$, see the discussion in section~\ref{sec:Methods/approximations/propagator}. As a check of our numerical results we have verified that taking $\mathscr{S} = 1/E_e^2$ for all diagrams, the vector component vanishes to very good precision. 

\subsection{Approximations for the intermediary propagators}

\begin{figure}
    \centering
    \resizebox{!}{\hsize}{\includegraphics{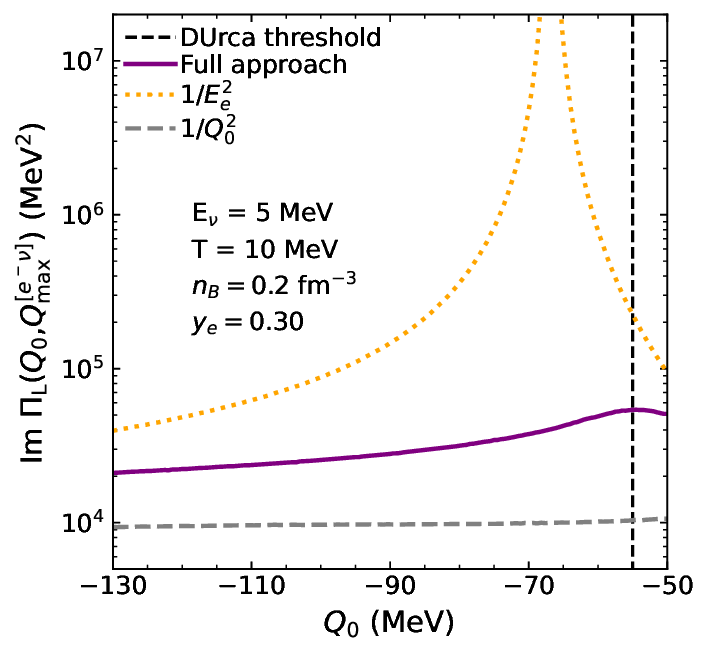}}
    \caption{Axial longitudinal component of the imaginary part of the polarization function as a function of the energy transfer $Q_0$. Results are presented for different choices for the intermediary propagators: our approach (Full), $\mathscr{S} = 1/E_e^2$ and $\mathscr{S} = 1/Q_0^2$. }
    \label{fig:denomin}
\end{figure}
As mentioned before, a common approximation used to simplify the calculation of the MUrca rates relies on choosing a simple form for the intermediary propagators, i.e. the function $\mathscr{S}$ as discussed in section~\ref{sec:Methods/approximations/propagator}. In our approach, later on referred to as the "full" approach, $\mathscr{S}$ in Eq.~\eqref{eq:intermediaryProp} includes the full dependence on energy and momentum of the intermediary particles. In addition we include in-medium effects on the mean-field level via effective masses and chemical potentials. The parameter $\Gamma$ is introduced to mimic in a very simple way a nucleonic quasi-particle life-time. In Ref.~\cite{Shternin2018}, the authors already put into light the role of the poles of these intermediate propagators in the enhancement of the MUrca rate when the kinematic conditions are close to the DUrca threshold, an effect that is missed within the common approximations. 

Let us now compare the results for the imaginary part of the hadronic polarization function within the different approaches. It is displayed as a function of energy exchange $Q_0$ in Fig.~\ref{fig:denomin} for a neutrino energy 5\;MeV, a temperature $T= 5$ MeV, a baryon number density $n_B = 0.2\;\mathrm{fm}^{-3}$ and an electron fraction $Y_e = 0.3$. Results for the "full" approach do show an enhancement of the results close to the DUrca threshold as discussed before. 
In many works, e.g. Refs.~\cite{Friman1979,Yakovlev1995, Yakovlev2001,Alford2018}, the denominator of $\mathscr{S}$ is reduced to the energy of the charged leptons squared. This case is represented in Fig.~\ref{fig:denomin} as a dotted orange line. Because the denominator of the intermediary propagator is reduced to the energy of the charged lepton, a divergence appears when it approaches zero. This is the case close to the thresholds for the different processes. Here we show the results as a function of the energy transfer $Q_0$. For the present  thermodynamic conditions and a neutrino energy $E_{\nu} =5$\;MeV, the electron capture opens at ${Q_{0; \rm ext}^{[e^- \bar{\nu}]} = -57.51\;{\rm MeV}}$ and the positron capture stops at ${Q_{0; \rm ext}^{[e^+ \bar{\nu}]]}= -58.49\;{\rm MeV}}$. Thus positron capture exists on the left hand side of the divergence, and the neutron decay on the right hand side.  The region in between these two values is not physical for any MUrca process. Clearly, close to the thresholds of the different processes, in this approach the result is largely overestimated compared with the full approach and remains for the present conditions dominant over a large range of $Q_0$ values. 

Another common approximation is to assume ${\mathscr{S} = 1/Q_0^2}$ as for instance applied in Ref.~\cite{Bacca2012}. In Fig.~\ref{fig:denomin} the corresponding results are shown as a black dashed line. For the present conditions the approach underestimates results by a factor of three. As for the approach with $\mathscr{S} = 1/E_e^2$, there is an unphysical divergence, this time around $Q_0 = 0$ which is, however, not visible in the range of values shown in the figure.

\subsection{Suppression of MUrca processes with respect to DUrca ones}\label{sec:Regimes}

It is in general assumed that MUrca type reactions are strongly suppressed with respect to DUrca type reactions when the latter is kinematically possible. In this section, we will review this question in the different regimes of temperature and density and corroborate our numerical results with a simple analytical estimation, which allows us to understand the reasons for the relative importance of MUrca and DUrca in the different regimes. 
\begin{figure*}
    \resizebox{\hsize}{!}{\includegraphics{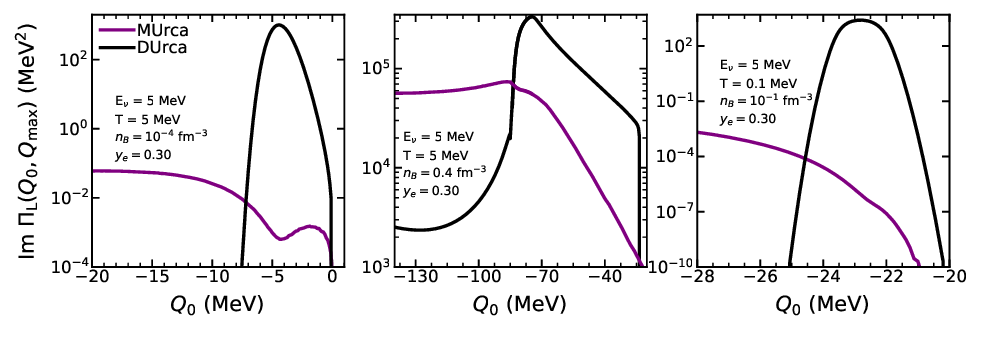}}
    \caption{Imaginary part of the hadronic polarization function as a function of the weak boson energy for the maximum value of the weak boson momentum norm. This quantity is represented for the modified Urca process and the direct Urca process.}
    \label{Fig:regime}
\end{figure*}
\subsubsection{Low temperature and high density regime}

For low temperatures ($T\lesssim 0.1$\;MeV) and high densities, that is to say in thermodynamic conditions relevant for cold neutron stars, the ratio between the MUrca and DUrca neutrino emission rates was thoroughly discussed in \textcite{Yakovlev2001}. The authors derive Urca neutrino rates at $\beta$-equilibrium in the Fermi surface approximation, with an intermediary propagator reduced to the inverse of the electron energy. In these conditions, they found a suppression of the MUrca process with respect to DUrca of six orders of magnitude.

In Fig.~\ref{Fig:regime} (right panel), we present results for the imaginary part of the hadronic polarization function for both the DUrca and MUrca process, for a low temperature $T=0.1$\;MeV, a relatively high baryon number density  of $n_B=0.1$\;fm$^{-3}$ typical of the values in the neutron star interior. The charged lepton fraction has been chosen $Y_e = 0.3$, a value, which is higher than typical values in a cold and $\beta$-equilibrated neutron star, however, different values of $Y_e$ lead to similar results. In accordance with \textcite{Yakovlev2001}, we find that MUrca is suppressed with respect to DUrca by six to eight orders of magnitude. Our calculations are performed with a full numerical computation, thus in particular \mo{we} use the full expression for the intermediate propagators and do not apply the Fermi surface approximation. The latter should of course be a very good approximation for these conditions. The reason for this strong suppression at low temperatures and high densities confirmed by our full calculation is the reduced phase space for MUrca processes, which becomes very pronounced at low temperature with the distribution functions approaching a Heaviside step function. 


\subsubsection{High and moderate temperature regime}

\begin{figure}
    \resizebox{\hsize}{!}{\includegraphics{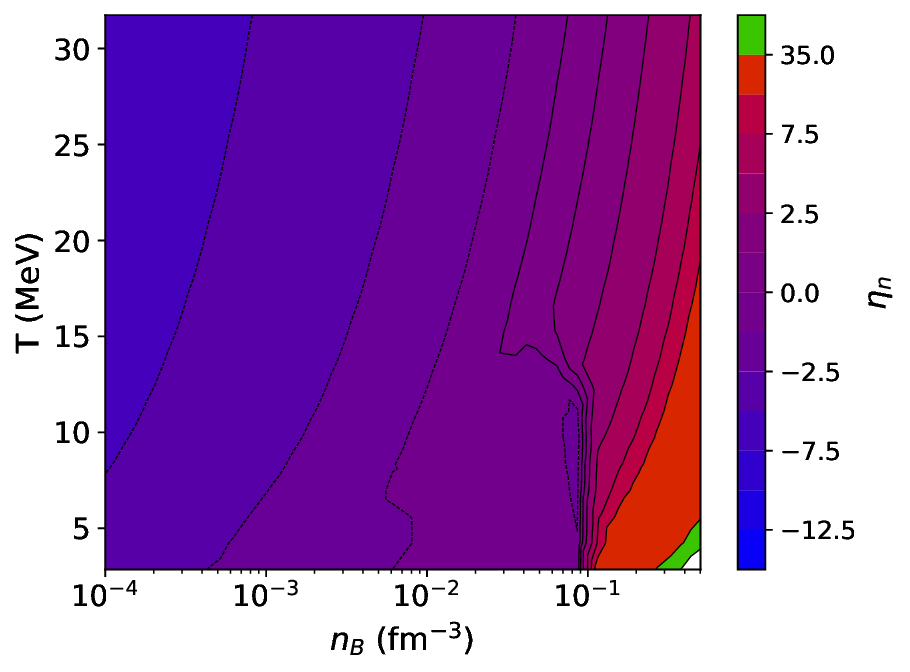}}
    \caption{Contour representation of the quantity ${\eta_n(T,n_B) = \big(\mu^*_n(T,n_B) - m^*_n(T,n_B)\big)/T}$ . }
    \label{fig:contourNeutrons}
\end{figure}

Before discussing the full numerical results, we present a simple approximation in this high and moderate temperature regime, which allows us to estimate the ratio between the MUrca and DUrca rates. 
While the DUrca process involves one incoming particle and one outgoing particle, the MUrca process involves two incoming and two outgoing particles, such that the integrands, denoted $I^{\rm Du}$ and $I^{\rm Mu}$ for DUrca and MUrca processes respectively, recover from phase space the following structure 
\begin{align}
    I^{\rm Mu} &\propto \frac{(m_1^*T)^{3/2}(m_2^*T)^{3/2}(m_3^*T)^{3/2}}{(2\pi^2)^3}\mathcal{F}_1 \mathcal{F}_3 (1-\mathcal{F}_2)(1-\mathcal{F}_4) \;, \nonumber \\ 
    I^{\rm Du} &\propto  \frac{(m_1^*T)^{3/2}}{2\pi^2} \mathcal{F}_1 (1-\mathcal{F}_2) \;.
    \label{eq:approxDuMu}
\end{align}
$m_i^*$ represents the effective mass of the nucleon $i$. The Fermi-Dirac distribution $\mathcal{F}_i$ of particle $i$ is given by 
\begin{equation}
    \mathcal{F}_i =  \frac{1}{1 + e^{\beta E_i - \eta_i}} \;,
\end{equation}
with $\eta_i = (\mu^*_i- m^*_i)/T$ and $\beta = 1/T$. 
For sufficiently hot and dilute matter, this distribution function can be approximated by its Maxwell-Boltzmann expression, 
\begin{equation}
    \mathcal{F}_i^{\rm Ht} = e^{-\beta E_i }e^{\eta_i} \;, \label{eq:apprixmationFermi}
\end{equation}
and be neglected in Eq.~\eqref{eq:approxDuMu} with respect to one (i.e. there is no Pauli blocking for outgoing nucleons). The ratio of the integrands then becomes 
\begin{equation}
    \frac{I^{\rm Mu}}{I^{\rm Du}} \propto e^{\eta_3}  \frac{(m^*_2T)^{3/2}(m^*_3 T)^{3/2} }{(2\pi^2)^2}e^{-\beta E_i } \;. \label{eq:factorPertub}
\end{equation}
To estimate the ratio of the rates, we can assume that the integration over the kinetic energy term with the matrix elements is of the same order for both MUrca and DUrca. Then, the factor containing $m_*^i T$ is of order $\mathcal{O}(1)$ such that the main factor determining the ratio of MUrca and DUrca rates in this regime is the exponential factor $e^{\eta_i}$, which gives us an indication of the order of magnitude of the ratio. To get an idea of the expected importance of MUrca with respect to DUrca, we thus represent the quantity $\eta_n$ for neutrons in Fig.~\ref{fig:contourNeutrons} in the temperature-baryon number density plane.  For the electron fraction we have chosen a typical value of $Y_e = 0.3$ and the EoS is SRO(APR); the qualitative conclusions are the same for a different value of the charged lepton fraction, for protons and for a different EoS. 

From the behavior of $\eta_n$ in Fig.~\ref{fig:contourNeutrons}, we can identify different regimes in temperature and baryon number density where we either expect MUrca processes to be of the same order or dominant with respect to DUrca, or to be clearly suppressed. We can conclude the following:
\begin{itemize}
    \item Along the contour lines for which $\eta$ is negative or very small, i.e. at very low densities, the MUrca processes are strongly suppressed relative to DUrca. Our rough estimate is confirmed by the full numerical results. As an example, we show the imaginary part of the hadronic polarization function for the thermodynamic conditions ${T=5\;{\rm MeV}}$, ${n_B = 10^{-4}\;{\rm fm}^{-3}}$ and ${Y_e = 0.3}$ in Fig.~\ref{Fig:regime} (left panel). The MUrca one is indeed suppressed by five orders of magnitude with respect to DUrca. In these conditions, it is reasonable to neglect the calculation of the MUrca processes upon DUrca being active. The reason is obviously that a process involving two incoming nucleons as for MUrca becomes increasingly difficult in dilute matter even if the distribution functions are subject to thermal broadening. 
    \item Around the zero contour line of $\eta$, the leading factor ${e^{\eta}}$ and therefore the ratio ${I^{\rm Mu} / I^{\rm Du}}$ is of order $\mathcal{O}(1)$. In that case, we expect both MUrca and DUrca rates to be of the same order of magnitude. This expectation is again confirmed by our numerical results, illustrated in Fig.~\ref{Fig:regime} (middle panel), where we show the hadronic polarization function for the thermodynamic conditions ${T=5\;{\rm MeV}}$, ${n_B=0.4}\;{\rm fm}^{-3}$ and ${Y_e=0.3}$. We thus find a regime at moderate temperatures and densities where due to the thermal effects in not too dilute matter, the MUrca rates are not necessarily suppressed with respect to DUrca and and it is important to compute the MUrca rates when DUrca is active.
\end{itemize}

\subsection{Neutrino and anti-neutrino absorption opacity}

\begin{figure}
    \resizebox{\hsize}{!}{\includegraphics{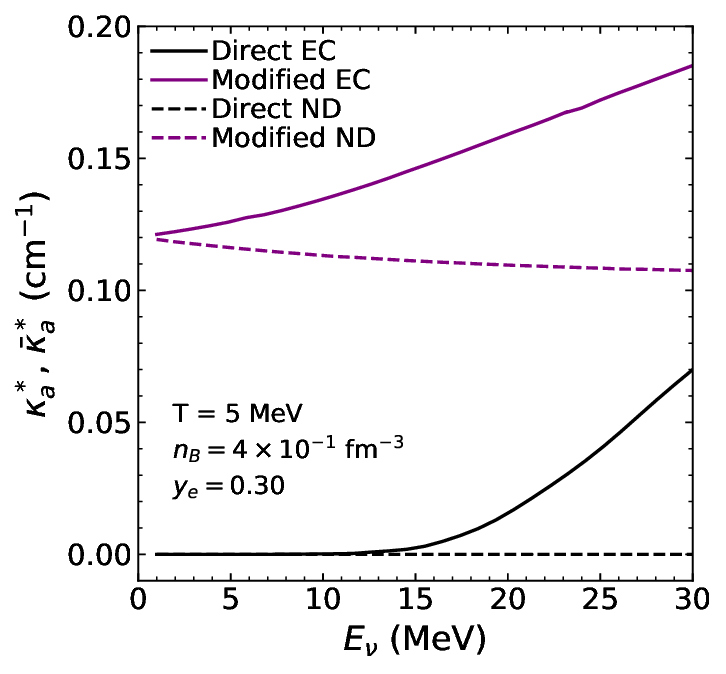}}
    \caption{Absorption opacity for direct and modified electron capture and neutron decays.}
    \label{Fig:leptons}
\end{figure}

\begin{figure}
    \resizebox{\hsize}{!}{\includegraphics{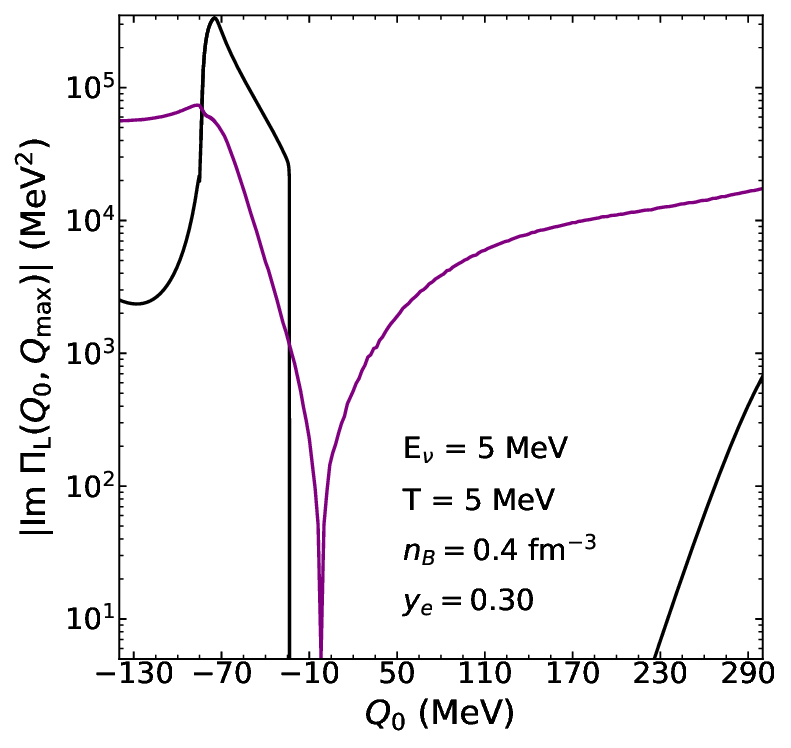}}
    \caption{Absolute value of the imaginary part of the hadronic polarization function as a function of the energy transfer for the maximum value of the norm of momentum transfer. This quantity is represented corresponding to MUrca reactions and DUrca ones, respectively,  for the thermodynamic conditions $T=5$\;MeV, ${n_B=0.4}$\;fm$^{-3}$ and $Y_e=0.3$.}
    \label{Fig:regimes2}
\end{figure}


Up to now we have only discussed results for the imaginary part of the hadronic polarization function associated either to MUrca or to DUrca type reactions. But of course, the final rates for the different processes do not depend only on $\Im \Pi$, but on the kinematic conditions for each reaction which determine the relevant range for energy and momentum transfer. At high density and low temperature as well as for dilute matter, where $\Im \Pi$ is suppressed for MUrca with respect to DUrca by several orders of magnitude upon the latter being active, the different energy and momentum transfer conditions will not change anything and the MUrca rates will be suppressed by several order of magnitude whenever DUrca reactions are possible. Then computing MUrca reaction rates is pertinent only for conditions where DUrca is kinematically forbidden or strongly suppressed as previously assumed in the literature. 

This is not the case in the moderate temperature and density regime identified above. Let us therefore discuss MUrca and DUrca opacities in this regime in more details now. The neutrino and anti-neutrino absorption opacities as function of the (anti-)neutrino energy are displayed in Fig.~\ref{Fig:leptons} for modified and direct electron captures and neutron decays for
the same thermodynamic conditions $T=5$\;MeV, ${n_B=0.4}$\;fm$^{-3}$ and $Y_e=0.3$ as before. From  Fig.~\ref{Fig:regime} (middle panel), where the imaginary part of the hadronic polarization function is shown for these thermodynamic conditions and a fixed value of the neutrino energy (5\;MeV), we had argued that the modified and direct processes should be of the same order. As can be seen, this is indeed the case for electron capture at not too small neutrino energies. However, the direct neutron decays are completely suppressed for these conditions. The modified neutron decay is not, and its absorption opacity remains more or less constant for the range of anti-neutrino energies shown. The direct electron capture is suppressed for values of the neutrino energy below $\approx$ 15\;MeV and then starts to increase. It is largely dominated by the modified electron capture opacity at low energies and becomes comparable to it for energies slightly above roughly 20 MeV. 

The behavior of the opacities in Fig.~\ref{Fig:leptons} can be understood by considering the energy transfers relevant for the corresponding process. For instance, for electron captures at $E_\nu = 5$ MeV, positive values of the energy transfer $Q_0$ largely contribute to the rate. We therefore redisplay the imaginary part of the hadronic polarization function as function of the energy transfer in Fig.~\ref{Fig:regimes2}, this time the absolute value on a logarithmic scale and for a larger range of energy transfers. It is obvious that for $Q_0 > 0$ the DUrca one is much smaller than in the case of MUrca, which explains the dominant opacity contribution from MUrca for this neutrino energy. Similar arguments can be applied to all other processes. 


\section{Summary and Conclusion}
In this work we have a performed for the first time a full numerical evaluation of rates for MUrca type processes in hot and dense matter. We have discussed the impact of several commonly used approximations on the computation of these rates. Our main findings are the following. (i) Contrary to what has been common practice in the literature, the vector contribution cannot be generally neglected. (ii) Approximating the propagators of intermediate nucleons by the charged lepton's energy or the energy transfer misses an enhancement close and above the DUrca threshold. We thereby confirm the results of \textcite{Shternin2018} and show that this enhancement is due to the opening of the DUrca channel for the intermediate propagators of the MUrca reactions. It is probably overestimated by the infinite nucleonic quasi-particle lifetime considered in Ref.~\cite{Shternin2018}. The effect is similar to the enhancement of the neutral current bremsstrahlung rate close to $Q_0 \to 0$ as already discussed in Ref.~\cite{Hannestad1998}. For our further numerical evaluations we have assumed a constant imaginary part for the nucleonic self-energy in the intermediate propagators, with a value of 10 MeV, adjusted such that MUrca rates far from the DUrca threshold are only marginally affected. A more thorough investigation of this point and in particular a consistent determination of the quasi-particle lifetime is left for future work. (iii) The Fermi surface approximation works very well at low temperatures, but thermal effects become important already at moderate temperatures. In particular we show that there is a regime at moderate temperatures and densities, where MUrca reactions are not necessarily suppressed with respect to DUrca ones when the latter process is active. This regime is potentially relevant to determine $\beta$-equilibrium conditions in binary neutron star merger remnants or the late cooling of a proto-neutron star. Our results show that a careful evaluation of MUrca rates is in order for these conditions. We have shown that this finding is mainly explained by phase space effects and thus do not expect a significant change in our conclusion upon improving on our approach, e.g. by going beyond the one-pion exchange approximation we have used. We should, however, be careful since the OPE is known to overestimate the matrix element. Another open question is here the importance of the quasi-particle lifetime in the intermediate propagators. A very short life-time could suppress the MUrca opacity, but for realistic values we do not expect an order of magnitude change. Thus we expect our findings to be robust, and our conclusion is that work should be dedicated to the MUrca reactions.

\begin{acknowledgments}
We thank A. Haber and M. Urban for insightful discussions, A. Sedrakian for useful comments on the manuscript, and T. Helpin for guidance with algebraic computing software. 
The authors acknowledge the financial support of the National Science Center, Poland grant 2018/29/B/ST9/02013 and 2019/33/B/ST9/00942, of the National Science Foundation grant Number PHY 21-16686, and the Agence Nationale de la Recherche (ANR) under contract ANR-22-CE31-0001-01. 
\end{acknowledgments}

\appendix

\section{Matsubara formalism for the hadronic polarization function}\label{Appendix:matsuFunction}

The imaginary part of the retarded hadronic polarization function is established in several steps. First we compute the relevant traces in spin and isospin space for each diagram. These traces imply a product of all the components described in section~\ref{sec:Methods/hadronFunction}: the nucleon propagators, the strong interaction matrix in OPE and the weak boson to nucleon vertex, for details see Appendix~\ref{Appendix:spin} and Appendix~\ref{Appendix:isospin}. Then we obtain an expression for the temperature polarization function in Matsubara formalism (see Ref.~\cite{Matsubara1955}). Since the zero component of the pion four-momenta has been neglected in our study (a common feature of the OPE approximation),  the sum over Matsubara frequencies involves only the denominators of nucleon propagators. It can be treated conveniently with the \textsc{MatsubaraSum} package for the \textsc{Mathematica} algebra software and analytic expressions are obtained. Then an analytic continuation can be performed to obtain the retarded function. 

Only the imaginary part of the retarded hadronic polarization function enters the rates. Since only the part obtained from the Matsubara sum has an imaginary part, the operation can be done easily. As mentioned before, we only keep the cuts related to MUrca type reactions and neglect the cuts contributing to corrections to the DUrca rates. The result of the above steps leads to the functions $\mathcal{M}$ presented in Eq.~\ref{eq:integral}, which are given by the following expressions for the different diagrams (diagram $V_1$ and $V_2$ are identical and we only list the expressions for $V_1$ in the following discussion)
\begin{widetext}
\begin{align}
    \Im \mathcal{M}_D &= \pi \frac{\mathcal{G}(\mathscr{E}_{\vec{p_1};x}, \mathscr{E}_{\vec{p_2};y}, \mathscr{E}_{\vec{p_3};z}, \mathscr{E}_{\vec{p_4};u})}{\big( \mathscr{E}_{\vec{p_6};w} +\mathscr{E}_{\vec{p_2};y} - \mathscr{E}_{\vec{p_3};z} - \mathscr{E}_{\vec{p_4};u} \big)^2}  \delta(\mathscr{E}_{\vec{p_1};x} + \mathscr{E}_{\vec{p_2};y} + Q_0 - \mathscr{E}_{\vec{p_3};z} -\mathscr{E}_{\vec{p_4};u}) \;, \label{MastuFunction1} \\
   \Im \mathcal{M}_E &= - \pi \frac{\mathcal{G}(\mathscr{E}_{\vec{p_1};x}, \mathscr{E}_{\vec{p_2};y}, \mathscr{E}_{\vec{p_3};z}, \mathscr{E}_{\vec{p_4};u})}{\big( \mathscr{E}_{\vec{p_6};w} +\mathscr{E}_{\vec{p_2};y} - \mathscr{E}_{\vec{p_3};z} - \mathscr{E}_{\vec{p_4};u} \big)^2} \delta(\mathscr{E}_{\vec{p_1};x} + \mathscr{E}_{\vec{p_2};y} - Q_0 - \mathscr{E}_{\vec{p_3};z} -\mathscr{E}_{\vec{p_4};u}) \label{MastuFunction2}  \\
     \Im \mathcal{M}_{V_1} & = \pi \frac{1}{(\mathscr{E}_{\vec{p_5};v} + \mathscr{E}_{\vec{p_2};y}-\mathscr{E}_{\vec{p_3};z}-\mathscr{E}_{\vec{p_4};u}) (\mathscr{E}_{\vec{p_1};x}+\mathscr{E}_{\vec{p_2};y}-\mathscr{E}_{\vec{p_4};u}-\mathscr{E}_{\vec{p_6};w})} \label{MastuFunction3} \\
    & \hspace{1.5cm} \times \bigg(  \mathcal{G}(\mathscr{E}_{\vec{p_1};x}, \mathscr{E}_{\vec{p_2};y}, \mathscr{E}_{\vec{p_3};z}, \mathscr{E}_{\vec{p_4};u}) \delta(\mathscr{E}_{\vec{p_1};x} + \mathscr{E}_{\vec{p_2};y} + Q_0 - \mathscr{E}_{\vec{p_3};z} -\mathscr{E}_{\vec{p_4};u} ) \nonumber  \\
    &\hspace{2.5cm}- \mathcal{G}(\mathscr{E}_{\vec{p_5};v}, \mathscr{E}_{\vec{p_2};y}, \mathscr{E}_{\vec{p_6};w}, \mathscr{E}_{\vec{p_4};u}) \delta(\mathscr{E}_{\vec{p_4};u}+\mathscr{E}_{\vec{p_6};w}+ Q_0 -\mathscr{E}_{\vec{p_2};y} - \mathscr{E}_{\vec{p_5};v}) \bigg) \;,  \nonumber \\
    \Im \mathcal{M}_{V_3} & = - \pi \frac{1}{(\mathscr{E}_{\vec{p_3};z}+\mathscr{E}_{\vec{p_4};u}-\mathscr{E}_{\vec{p_1};x}-\mathscr{E}_{\vec{p_6};w}) (\mathscr{E}_{\vec{p_5};v}+\mathscr{E}_{\vec{p_2};y}-\mathscr{E}_{\vec{p_3};z}-\mathscr{E}_{\vec{p_4};u})} \label{MastuFunction4} \\
    &\hspace{1.5cm} \times \bigg( \mathcal{G}(\mathscr{E}_{\vec{p_5};v}, \mathscr{E}_{\vec{p_6};w}, \mathscr{E}_{\vec{p_3};z}, \mathscr{E}_{\vec{p_4};u}) \delta(\mathscr{E}_{\vec{p_6};w} + \mathscr{E}_{\vec{p_5};v} -Q_0  -\mathscr{E}_{\vec{p_3};z} - \mathscr{E}_{\vec{p_4};u} ) \nonumber \\
    & \hspace{2.5cm} - \mathcal{G}(\mathscr{E}_{\vec{p_1};x}, \mathscr{E}_{\vec{p_2};y}, \mathscr{E}_{\vec{p_3};z}, \mathscr{E}_{\vec{p_4};u}) \delta(\mathscr{E}_{\vec{p_1};x}+ \mathscr{E}_{\vec{p_2};y} +Q_0 -\mathscr{E}_{\vec{p_3};z} -\mathscr{E}_{\vec{p_4};u})\bigg) \;. \nonumber 
\end{align}
\end{widetext}
The indices $\{x,y,z,u,v,w\}$ refer to the isospin nature of the nucleons $(1)$, $(2)$, $(3)$, $(4)$, $(5)$ and $(6)$ respectively, and $\vec{p}_1$, $\vec{p}_2$, $\vec{p}_3$, $\vec{p}_4$, $\vec{p}_5$ and $\vec{p}_6$ label the nucleon momenta. Conservation of three-momenta for the hadron part of the MUrca process is presented in Table~\ref{tab:momentConservation}. The function $\mathcal{G}$ is expressed in terms of the nucleonic Fermi-Dirac distribution functions $\mathcal{F}_F$ as 
\begin{align}
    \mathcal{G}(\alpha, \beta, \gamma, \delta) &= \mathcal{F}_F(\alpha) \mathcal{F}_F(\beta) \big(1- \mathcal{F}_F(\gamma) \big) \big(1-\mathcal{F}_F(\delta)\big) \nonumber \\
    & -  \big(1-\mathcal{F}_F(\alpha)  \big) \big(1- \mathcal{F}_F(\beta)\big) \mathcal{F}_F(\gamma)\mathcal{F}_F(\delta) \;. \nonumber 
\end{align}
and clearly shows the structure of the reactions with two incoming and two outgoing nucleons.

\begin{table}
     \centering
     \begin{tabular}{|c|c|c|c|c|}
     \hline
          Nucleon & $D$ & $E$ & $V_1$ & $V_3$ \\ \hline
          $(5)$ & $\vec{p}_1 + \vec{Q}$ &  $\vec{p}_4-\vec{Q}$ & $\vec{p}_1+\vec{Q}$ & $\vec{p}_1+\vec{Q}$ \\
          $(6)$ & $\vec{p}_1+\vec{Q}$ &  $\vec{p}_4-\vec{Q}$ & $\vec{p}_3 - \vec{Q}$ & $\vec{p}_2+\vec{Q}$ \\
          \hline
     \end{tabular}
     \caption{Momenta of the intermediate nucleons via three-momentum conservation for the self-energy corrections $D$ and $E$ and the vertex corrections $V_1$ and $V_3$.}
     \label{tab:momentConservation}
 \end{table}

\section{Trace in isospin space} \label{Appendix:isospin}

Among the different elements of the hadronic polarization function presented in section~\ref{sec:Methods/hadronFunction}, only the weak boson to nucleon vertex and the nucleon propagators contains terms to be traced in the isospin space. Operating the trace and summing over all the possible isospin natures of nucleons in play gives us the expression of $I(X)$, with $X$ referring to the diagram. For example, the explicit formulation for one of the sub-diagrams (ways nucleons can be exchanged in the diagrams) of the first self-energy contribution, here denoted $I(D_a)$, is given by

\small{
\begin{align}
     I(D_a) &=  \frac{1}{64} \sum_{\{t_x,t_y,t_z,t_u,t_v,t_w\}}  \Tr[(1+t_y\tau^3)\tau_i (1+t_z\tau^3)\tau_j] \nonumber \\
    & \Tr[\tau^{+}(1+t_{x}\tau^3)\tau^{-} (1+t_v\tau^3)\tau^i (1+t_{u}\tau^3)\tau^{j}(1+t_{w}\tau^3)] \nonumber \\ 
    & \times  S_x(p_1) S_y(p_2) S_z(p_3)  S_u(p_4) S_v(p_5) S_w(p_6) \nonumber \\
    &= S_n(p_6)^2 \Big[ S_n(p_2) S_n(p_3) S_n(p_4) \nonumber \\
    & +  S_p(p_2) \big( 4  S_n(p_3) S_p(p_4) + S_p(p_3) S_n(p_4)\big) \Big] S_p(p_1) \;. \nonumber 
\end{align}}
For each diagram, we can present the authorized isospin combinations as a vector whose components correspond to the different sub-diagrams $(a \dots d)$. In the following, the notation $(xyzuvw)$ designates the isospin nature of nucleons $(1)$,  $(2)$,  $(3)$, $(4)$, $(5)$ and $(6)$ presented in the diagrams of Fig.~\ref{fig:murcadiagrams}, in that order
\begin{align}
    &I(D) = \begin{pmatrix} 
                (pnnnnn) + 4(ppnpnn) + (pppnnn) \\ (pnnnnn) + 4(pppnnn) + (ppnpnn) \\ (pnnnnn) -2(ppnpnn) -2(pppnnn) \\ (pnnnnn) -2(ppnpnn) -2(pppnnn)
            \end{pmatrix} \;, \\
    &I(E) = \begin{pmatrix} 
                (pppnpp) + 4(npnnpp) + (pnnnpp) \\ (pppnpp) + 4(pnnnpp) + (npnnpp) \\(pppnpp) -2(npnnpp) -2(pnnnpp) \\(pppnpp) -2(pnnnpp) -2(npnnpp) 
            \end{pmatrix} \;, \\
    &I(V_1) = \begin{pmatrix} 
                 2(ppnpnp) + 2(pnnnnp) \\ -(pnnnnp) -(ppnpnp) \\2(pnnnnp) -(ppnpnp) \\ -(pnnnnp) + 2(ppnpnp)
            \end{pmatrix} \;, \\
    &I(V_3) = \begin{pmatrix} 
                 -2(ppnpnn) -2(pppnnn)\\ -2(ppnpnn) -2(pppnnn) \\ 4(pppnnn) + (ppnpnn) \\4(ppnpnn) + (pppnnn) 
            \end{pmatrix} \;, 
\end{align}
It is interesting to note that $I(X)$ can be deduced from Fig.~\ref{fig:murcadiagrams}, by respecting the isospin conservation in the diagrams.

\section{Trace in spin space}\label{Appendix:spin}

The strong interaction matrix in the OPE approximation and the weak boson to nucleon vertex enter the trace in spin space. As introduced in Eq.~\eqref{eq:integral}, $\mathcal{X}_{Y; \rm spin}^{\alpha \beta} (\vec{k},\vec{k}') $ depends on the momenta $\vec{k}$ and $\vec{k}'$ of the exchanged pions. We understand that greek indices are $\in [0,1,2,3]$ and latin indices refer to space components $\in [1,2,3]$. Let us introduce the following notation for the pion momenta in terms of the momenta of the involved nucleons
${\vec{k}_1 =\vec{p}_3-\vec{p}_2}$ and ${\vec{k}_2 =\vec{p}_4-\vec{p}_2}$,
 ${\vec{k}_3 =\vec{p}_1-\vec{p}_3 }$,
 ${\vec{k}_4 = \vec{p}_4-\vec{p}_1 }$, 
as well as the following functions  
\begin{align*}
N(\vec{k},\vec{k'}) & = \frac{1}{(|\vec{k}|^2+m_{\pi}^2)(|\vec{k}'|^2+m_{\pi}^2)} \;,\\
    f_1(\vec{k},\vec{k}') &= 4(\vec{k} \cdot \vec{k}')^2 \;, \\
    f_2(\vec{k},\vec{k}') &= 2 \Big(2 (\vec{k}\cdot \vec{k}')^2-|\vec{k}|^2|\vec{k}'|^2\Big)\;, \\
    f_3(\vec{k},\vec{k'}) &= 4 \,\vec{k}^2 \;\vec{k'}^2 \; , \\
    h^j(\vec{k}, \vec{k}') &= 4 {\rm i} \,(\vec{k} \cdot \vec{k}')\, ( \vec{k} \wedge \vec{k}')^j \;, \\
    g_1^{ij}(\vec{k},\vec{k}') &= 4 (\vec{k}\wedge \vec{k}')^i \,(\vec{k} \wedge \vec{k}')^j \;,\\
    g_2^{ij}(\vec{k},\vec{k'}) &= 4 (\vec{k}\cdot\vec{k}') \, k^i (k')^j\;, \\
    g_3^{ij}(\vec{k},\vec{k'}) &= 4 |\vec{k}|^2 \, (k')^i (k')^j\;.
\end{align*}

The spin traces then lead to the following terms for the self-energy diagram D~\footnote{Since the mixed terms proportional to $c_A c_V$ vanish upon summing up the different diagrams and contracting with the lepton tensor, we do not list them here and put the corresponding elements to zero.} 
\begin{widetext}
\begin{align}
   \mathcal{X}_{D,a}^{\alpha \beta} (\vec{k}_1) &= \bigg(\frac{f_{\pi NN}}{m_{\pi}}\bigg)^4 \, N(\vec{k}_1,\vec{k}_1)  \begin{pmatrix} f_1(\vec{k}_1, \vec{k}_1) c_V^2 & 0 \\ 0 &  f_1(\vec{k}_1,\vec{k}_1) \delta^{ij} c_A^2 \end{pmatrix} \;,\\
   \mathcal{X}_{D,b}^{\alpha \beta} (\vec{k}_2) &= \bigg(\frac{f_{\pi NN}}{m_{\pi}}\bigg)^4 \, N(\vec{k}_2,\vec{k}_2)  \begin{pmatrix} f_1(\vec{k}_2, \vec{k}_2) c_V^2 & 0 \\ 0 &  f_1(\vec{k}_2,\vec{k}_2) \delta^{ij} c_A^2 \end{pmatrix}\;, \\
    \mathcal{X}_{D,c}^{\alpha \beta} (\vec{k}_1, \vec{k}_2) &= -\bigg(\frac{f_{\pi NN}}{m_{\pi}}\bigg)^4 N(\vec{k}_1,\vec{k}_2) \begin{pmatrix} f_2(\vec{k}_1, \vec{k}_2) c_V^2 & 0
    \\ 0
    & f_2(\vec{k}_1, \vec{k}_2)\delta^{ij} c_A^2 \end{pmatrix} \, \\
    \mathcal{X}_{D,d}^{\alpha \beta} (\vec{k}_1, \vec{k}_2) &= -\bigg(\frac{f_{\pi NN}}{m_{\pi}}\bigg)^4 N(\vec{k}_1,\vec{k}_2) \begin{pmatrix} f_2(\vec{k}_1, \vec{k}_2) c_V^2 & 0
    \\ 0  
    & f_2(\vec{k}_1, \vec{k}_2)\delta^{ij} c_A^2 \end{pmatrix}\;.
\end{align}
\end{widetext}

The spin trace for diagram $E$ leads to the same terms as diagram D except that for the pion momenta 
$\vec{k_2} \to \vec{k_3}$.

The spin trace for $V_1$ gives the following terms 
\begin{widetext}
\begin{align}
   \mathcal{X}_{V_1,a}^{\alpha \beta} (\vec{k}_1, \vec{k}_4) &=  \bigg(\frac{f_{\pi NN}}{m_{\pi}}\bigg)^4 \,N(\vec{k}_1,\vec{k}_4)\begin{pmatrix} f_1(\vec{k}_1, \vec{k}_4) c_V^2 & 0
   \\ 
   0 & 
   g_1^{ij}(\vec{k}_1,\vec{k}_4) c_A^2
   \end{pmatrix} \;, \\
    \mathcal{X}_{V_1,b}^{\alpha \beta} (\vec{k}_2) &=  \bigg(\frac{f_{\pi NN}}{m_{\pi}}\bigg)^4\,N(\vec{k}_2,\vec{k}_2) \begin{pmatrix} f_1(\vec{k}_2, \vec{k}_2) c_V^2 &  0
    \\ 
    0 & \big[ -f_1(\vec{k}_2,\vec{k}_2)\delta^{ij} + 2 g_2^{ij}(\vec{k}_2,\vec{k}_2) \big] c_A^2 
    \end{pmatrix} \;, \\
     \mathcal{X}_{V_1,c}^{\alpha \beta} (\vec{k}_2,\vec{k}_4) &=  - \bigg(\frac{f_{\pi NN}}{m_{\pi}}\bigg)^4 \, N(\vec{k}_2,\vec{k}_4)\begin{pmatrix} f_2(\vec{k}_2, \vec{k}_4) c_V^2 & 0
     \\ 0 
     & \big[- f_2(\vec{k}_2,\vec{k}_4) \delta^{ij} + g_2^{ij}(\vec{k}_2,\vec{k}_4) + g_2 ^{ji}(\vec{k}_2,\vec{k}_4) - g_3^{ij}(\vec{k}_4,\vec{k}_2)  ]c_A^2
     \end{pmatrix} \;, \\
     \mathcal{X}_{V_1,d}^{\alpha \beta} (\vec{k}_1,\vec{k}_2) &= -\bigg(\frac{f_{\pi NN}}{m_{\pi}}\bigg)^4 \, N(\vec{k}_1,\vec{k}_2)\begin{pmatrix} f_2(\vec{k}_1, \vec{k}_2) c_V^2 & 0
     \\ 0  
     & \big[- f_2(\vec{k}_1,\vec{k}_2) \delta^{ij} + g_2^{ij}(\vec{k}_1,\vec{k}_2) + g_2 ^{ji}(\vec{k}_1,\vec{k}_2) - g_3^{ij}(\vec{k}_1,\vec{k}_2)  ]c_A^2
     \end{pmatrix} \;,
\end{align}
\end{widetext}

and for diagram $V_3$ the following terms are obtained from the spin trace 
\begin{widetext}
\begin{align} 
     \mathcal{X}_{V_3,a}^{\alpha \beta} (\vec{k}_2, \vec{k}_3) &=  \bigg(\frac{f_{\pi NN}}{m_{\pi}}\bigg)^4 \, N(\vec{k}_2,\vec{k}_3)\begin{pmatrix} f_1(\vec{k}_2, \vec{k}_3) c_V^2 & 0 
     \\ 0
     &  - g_1^{ij}(\vec{k}_2,\vec{k}_3) c_A^2
     \end{pmatrix} \;, \\
      \mathcal{X}_{V_3,b}^{\alpha \beta} (\vec{k}_1, \vec{k}_4) &=  \bigg(\frac{f_{\pi NN}}{m_{\pi}}\bigg)^4 \, N(\vec{k}_1,\vec{k}_4)\begin{pmatrix} f_1(\vec{k}_1, \vec{k}_4) c_V^2 & 0
      \\ 0
      & -g_1^{ij}(\vec{k}_1,\vec{k}_4) c_A^2
      \end{pmatrix} \;, \\
      \mathcal{X}_{V_3,c}^{\alpha \beta} (\vec{k}_2,\vec{k}_4) &=- \bigg(\frac{f_{\pi NN}}{m_{\pi}}\bigg)^4 \, N(\vec{k}_2,\vec{k}_4) \begin{pmatrix} f_2(\vec{k}_2, \vec{k}_4) c_V^2 & 0
      \\0
      & \big[-g_1^{ij}(\vec{k}_2,\vec{k}_4) + \delta^{ij}\, 
      \frac{f_3(\vec{k}_2,\vec{k}_4)}{2}
      ] c_A^2
      \end{pmatrix} \;, \\
      \mathcal{X}_{V_3,d}^{\alpha \beta} (\vec{k}_1,\vec{k}_3) &= - \bigg(\frac{f_{\pi NN}}{m_{\pi}}\bigg)^4 \, N(\vec{k}_1,\vec{k}_3)\begin{pmatrix} f_2(\vec{k}_1, \vec{k}_3) c_V^2 & 0
      \\ 0
      & \big[-g_1^{ij}(\vec{k}_1,\vec{k}_3) + \delta^{ij}\, 
      \frac{f_3(\vec{k}_1,\vec{k}_3)}{2}
      ] c_A^2
    \end{pmatrix} \;. 
\end{align}
\end{widetext}


\section{Numerical integration}\label{Appendix:MC}

The calculation of the MUrca neutrino rates involves in addition an integral over phase space which is performed numerically. The integral which is required to compute the hadronic part of the process is eight dimensional and the lepton part adds two more dimensions. The numerical cost of this multi-dimensional integration is one of the reasons why approximations are often taken to simplify the calculation and obtain an analytical expression of the MUrca rates. In this paper, we do not provide analytical expressions for the MUrca process to avoid approximations, and provide results based on numerical integration. 

The Monte-Carlo approach \cite{Metropolis1953} is particularly well suited to operate highly dimensional integrals. We use this method to compute the eight dimensional integral related to the hadronic part of the MUrca processes. In practice, an importance sampling Monte-Carlo method is used with the Fermi-Dirac distribution functions of the incoming nucleons as weights for the integration over energies and a uniform distribution is used to operate the angular integration in $\phi$ and $\cos\theta$ in spherical coordinates. The figures presented in this paper are based on calculations with $10^5$ to $10^6$ points. Note that the Fermi-Dirac distribution at low temperatures resembles a Heaviside function, such that a large number of points lead to a zero contribution in the integral, compared to high temperature calculations in which temperature broadening applies. Therefore, at fixed accuracy, more points are required at low temperatures compared to high temperatures. 

The remaining integration over the leptonic phase space for obtaining the opacities is done with a two-dimensional Gauss-Legendre quadrature.


\begin{thebibliography}{65}%
\makeatletter
\providecommand \@ifxundefined [1]{%
 \@ifx{#1\undefined}
}%
\providecommand \@ifnum [1]{%
 \ifnum #1\expandafter \@firstoftwo
 \else \expandafter \@secondoftwo
 \fi
}%
\providecommand \@ifx [1]{%
 \ifx #1\expandafter \@firstoftwo
 \else \expandafter \@secondoftwo
 \fi
}%
\providecommand \natexlab [1]{#1}%
\providecommand \enquote  [1]{``#1''}%
\providecommand \bibnamefont  [1]{#1}%
\providecommand \bibfnamefont [1]{#1}%
\providecommand \citenamefont [1]{#1}%
\providecommand \href@noop [0]{\@secondoftwo}%
\providecommand \href [0]{\begingroup \@sanitize@url \@href}%
\providecommand \@href[1]{\@@startlink{#1}\@@href}%
\providecommand \@@href[1]{\endgroup#1\@@endlink}%
\providecommand \@sanitize@url [0]{\catcode `\\12\catcode `\$12\catcode
  `\&12\catcode `\#12\catcode `\^12\catcode `\_12\catcode `\%12\relax}%
\providecommand \@@startlink[1]{}%
\providecommand \@@endlink[0]{}%
\providecommand \url  [0]{\begingroup\@sanitize@url \@url }%
\providecommand \@url [1]{\endgroup\@href {#1}{\urlprefix }}%
\providecommand \urlprefix  [0]{URL }%
\providecommand \Eprint [0]{\href }%
\providecommand \doibase [0]{http://dx.doi.org/}%
\providecommand \selectlanguage [0]{\@gobble}%
\providecommand \bibinfo  [0]{\@secondoftwo}%
\providecommand \bibfield  [0]{\@secondoftwo}%
\providecommand \translation [1]{[#1]}%
\providecommand \BibitemOpen [0]{}%
\providecommand \bibitemStop [0]{}%
\providecommand \bibitemNoStop [0]{.\EOS\space}%
\providecommand \EOS [0]{\spacefactor3000\relax}%
\providecommand \BibitemShut  [1]{\csname bibitem#1\endcsname}%
\let\auto@bib@innerbib\@empty
\bibitem [{\citenamefont {Prakash}\ \emph {et~al.}(1997)\citenamefont
  {Prakash}, \citenamefont {Bombaci}, \citenamefont {Prakash}, \citenamefont
  {Ellis}, \citenamefont {Lattimer},\ and\ \citenamefont
  {Knorren}}]{Prakash:1996xs}%
  \BibitemOpen
  \bibfield  {author} {\bibinfo {author} {\bibfnamefont {M.}~\bibnamefont
  {Prakash}}, \bibinfo {author} {\bibfnamefont {I.}~\bibnamefont {Bombaci}},
  \bibinfo {author} {\bibfnamefont {M.}~\bibnamefont {Prakash}}, \bibinfo
  {author} {\bibfnamefont {P.~J.}\ \bibnamefont {Ellis}}, \bibinfo {author}
  {\bibfnamefont {J.~M.}\ \bibnamefont {Lattimer}}, \ and\ \bibinfo {author}
  {\bibfnamefont {R.}~\bibnamefont {Knorren}},\ }\href {\doibase
  10.1016/S0370-1573(96)00023-3} {\bibfield  {journal} {\bibinfo  {journal}
  {Phys. Rept.}\ }\textbf {\bibinfo {volume} {280}},\ \bibinfo {pages} {1}
  (\bibinfo {year} {1997})},\ \Eprint {http://arxiv.org/abs/nucl-th/9603042}
  {arXiv:nucl-th/9603042} \BibitemShut {NoStop}%
\bibitem [{\citenamefont {Oertel}\ \emph {et~al.}(2017)\citenamefont {Oertel},
  \citenamefont {Hempel}, \citenamefont {Kl\"ahn},\ and\ \citenamefont
  {Typel}}]{Oertel:2016bki}%
  \BibitemOpen
  \bibfield  {author} {\bibinfo {author} {\bibfnamefont {M.}~\bibnamefont
  {Oertel}}, \bibinfo {author} {\bibfnamefont {M.}~\bibnamefont {Hempel}},
  \bibinfo {author} {\bibfnamefont {T.}~\bibnamefont {Kl\"ahn}}, \ and\
  \bibinfo {author} {\bibfnamefont {S.}~\bibnamefont {Typel}},\ }\href
  {\doibase 10.1103/RevModPhys.89.015007} {\bibfield  {journal} {\bibinfo
  {journal} {Rev. Mod. Phys.}\ }\textbf {\bibinfo {volume} {89}},\ \bibinfo
  {pages} {015007} (\bibinfo {year} {2017})},\ \Eprint
  {http://arxiv.org/abs/1610.03361} {arXiv:1610.03361 [astro-ph.HE]}
  \BibitemShut {NoStop}%
\bibitem [{\citenamefont {Burgio}\ and\ \citenamefont
  {Fantina}(2018)}]{FiorellaBurgio:2018dga}%
  \BibitemOpen
  \bibfield  {author} {\bibinfo {author} {\bibfnamefont {F.~G.}\ \bibnamefont
  {Burgio}}\ and\ \bibinfo {author} {\bibfnamefont {A.~F.}\ \bibnamefont
  {Fantina}},\ }\href {\doibase 10.1007/978-3-319-97616-7_6} {\bibfield
  {journal} {\bibinfo  {journal} {Astrophys. Space Sci. Libr.}\ }\textbf
  {\bibinfo {volume} {457}},\ \bibinfo {pages} {255} (\bibinfo {year}
  {2018})},\ \Eprint {http://arxiv.org/abs/1804.03020} {arXiv:1804.03020
  [nucl-th]} \BibitemShut {NoStop}%
\bibitem [{\citenamefont {{Stephenson}}\ and\ \citenamefont
  {{Green}}(2002)}]{Stephenson2002}%
  \BibitemOpen
  \bibfield  {author} {\bibinfo {author} {\bibfnamefont {F.~R.}\ \bibnamefont
  {{Stephenson}}}\ and\ \bibinfo {author} {\bibfnamefont {D.~A.}\ \bibnamefont
  {{Green}}},\ }\href@noop {} {\bibfield  {journal} {\bibinfo  {journal}
  {Historical supernovae and their remnants}\ }\textbf {\bibinfo {volume} {5}}
  (\bibinfo {year} {2002})}\BibitemShut {NoStop}%
\bibitem [{\citenamefont {{Janka}}(2012)}]{Janka2012}%
  \BibitemOpen
  \bibfield  {author} {\bibinfo {author} {\bibfnamefont {H.-T.}\ \bibnamefont
  {{Janka}}},\ }\href {\doibase 10.1146/annurev-nucl-102711-094901} {\bibfield
  {journal} {\bibinfo  {journal} {Annual Review of Nuclear and Particle
  Science}\ }\textbf {\bibinfo {volume} {62}},\ \bibinfo {pages} {407}
  (\bibinfo {year} {2012})},\ \Eprint {http://arxiv.org/abs/1206.2503}
  {arXiv:1206.2503 [astro-ph.SR]} \BibitemShut {NoStop}%
\bibitem [{\citenamefont {{Burrows}}\ and\ \citenamefont
  {{Lattimer}}(1987)}]{Burrows1987}%
  \BibitemOpen
  \bibfield  {author} {\bibinfo {author} {\bibfnamefont {A.}~\bibnamefont
  {{Burrows}}}\ and\ \bibinfo {author} {\bibfnamefont {J.~M.}\ \bibnamefont
  {{Lattimer}}},\ }\href {\doibase 10.1086/184938} {\bibfield  {journal}
  {\bibinfo  {journal} {Astrophysical Journal Letters}\ }\textbf {\bibinfo
  {volume} {318}},\ \bibinfo {pages} {L63} (\bibinfo {year}
  {1987})}\BibitemShut {NoStop}%
\bibitem [{\citenamefont {Mirizzi}\ \emph {et~al.}(2016)\citenamefont
  {Mirizzi}, \citenamefont {Tamborra}, \citenamefont {Janka}, \citenamefont
  {Saviano}, \citenamefont {Scholberg}, \citenamefont {Bollig}, \citenamefont
  {H{\"u}depohl},\ and\ \citenamefont {Chakraborty}}]{Mirizzi:2015eza}%
  \BibitemOpen
  \bibfield  {author} {\bibinfo {author} {\bibfnamefont {A.}~\bibnamefont
  {Mirizzi}}, \bibinfo {author} {\bibfnamefont {I.}~\bibnamefont {Tamborra}},
  \bibinfo {author} {\bibfnamefont {H.-T.}\ \bibnamefont {Janka}}, \bibinfo
  {author} {\bibfnamefont {N.}~\bibnamefont {Saviano}}, \bibinfo {author}
  {\bibfnamefont {K.}~\bibnamefont {Scholberg}}, \bibinfo {author}
  {\bibfnamefont {R.}~\bibnamefont {Bollig}}, \bibinfo {author} {\bibfnamefont
  {L.}~\bibnamefont {H{\"u}depohl}}, \ and\ \bibinfo {author} {\bibfnamefont
  {S.}~\bibnamefont {Chakraborty}},\ }\href {\doibase
  10.1393/ncr/i2016-10120-8} {\bibfield  {journal} {\bibinfo  {journal} {Riv.
  Nuovo Cim.}\ }\textbf {\bibinfo {volume} {39}},\ \bibinfo {pages} {1}
  (\bibinfo {year} {2016})},\ \Eprint {http://arxiv.org/abs/1508.00785}
  {arXiv:1508.00785 [astro-ph.HE]} \BibitemShut {NoStop}%
\bibitem [{\citenamefont {Roberts}(2012)}]{Roberts:2012zza}%
  \BibitemOpen
  \bibfield  {author} {\bibinfo {author} {\bibfnamefont {L.~F.}\ \bibnamefont
  {Roberts}},\ }\href {\doibase 10.1088/0004-637X/755/2/126} {\bibfield
  {journal} {\bibinfo  {journal} {Astrophys. J.}\ }\textbf {\bibinfo {volume}
  {755}},\ \bibinfo {pages} {126} (\bibinfo {year} {2012})},\ \Eprint
  {http://arxiv.org/abs/1205.3228} {arXiv:1205.3228 [astro-ph.HE]} \BibitemShut
  {NoStop}%
\bibitem [{\citenamefont {{Pascal}}\ \emph {et~al.}(2022)\citenamefont
  {{Pascal}}, \citenamefont {{Novak}},\ and\ \citenamefont
  {{Oertel}}}]{Pascal2022}%
  \BibitemOpen
  \bibfield  {author} {\bibinfo {author} {\bibfnamefont {A.}~\bibnamefont
  {{Pascal}}}, \bibinfo {author} {\bibfnamefont {J.}~\bibnamefont {{Novak}}}, \
  and\ \bibinfo {author} {\bibfnamefont {M.}~\bibnamefont {{Oertel}}},\ }\href
  {\doibase 10.1093/mnras/stac016} {\bibfield  {journal} {\bibinfo  {journal}
  {Monthly Notices of the Royal Astronomical Society}\ }\textbf {\bibinfo
  {volume} {511}},\ \bibinfo {pages} {356} (\bibinfo {year} {2022})},\ \Eprint
  {http://arxiv.org/abs/2201.01955} {arXiv:2201.01955 [nucl-th]} \BibitemShut
  {NoStop}%
\bibitem [{\citenamefont {{Endrizzi}}\ \emph {et~al.}(2020)\citenamefont
  {{Endrizzi}}, \citenamefont {{Perego}}, \citenamefont {{Fabbri}},
  \citenamefont {{Branca}}, \citenamefont {{Radice}}, \citenamefont
  {{Bernuzzi}}, \citenamefont {{Giacomazzo}}, \citenamefont {{Pederiva}},\ and\
  \citenamefont {{Lovato}}}]{Endrizzi2020}%
  \BibitemOpen
  \bibfield  {author} {\bibinfo {author} {\bibfnamefont {A.}~\bibnamefont
  {{Endrizzi}}}, \bibinfo {author} {\bibfnamefont {A.}~\bibnamefont
  {{Perego}}}, \bibinfo {author} {\bibfnamefont {F.~M.}\ \bibnamefont
  {{Fabbri}}}, \bibinfo {author} {\bibfnamefont {L.}~\bibnamefont {{Branca}}},
  \bibinfo {author} {\bibfnamefont {D.}~\bibnamefont {{Radice}}}, \bibinfo
  {author} {\bibfnamefont {S.}~\bibnamefont {{Bernuzzi}}}, \bibinfo {author}
  {\bibfnamefont {B.}~\bibnamefont {{Giacomazzo}}}, \bibinfo {author}
  {\bibfnamefont {F.}~\bibnamefont {{Pederiva}}}, \ and\ \bibinfo {author}
  {\bibfnamefont {A.}~\bibnamefont {{Lovato}}},\ }\href {\doibase
  10.1140/epja/s10050-019-00018-6} {\bibfield  {journal} {\bibinfo  {journal}
  {European Physical Journal A}\ }\textbf {\bibinfo {volume} {56}},\ \bibinfo
  {eid} {15} (\bibinfo {year} {2020})},\ \Eprint
  {http://arxiv.org/abs/1908.04952} {arXiv:1908.04952 [astro-ph.HE]}
  \BibitemShut {NoStop}%
\bibitem [{\citenamefont {{Foucart}}(2023)}]{Foucart2023}%
  \BibitemOpen
  \bibfield  {author} {\bibinfo {author} {\bibfnamefont {F.}~\bibnamefont
  {{Foucart}}},\ }\href {\doibase 10.1007/s41115-023-00016-y} {\bibfield
  {journal} {\bibinfo  {journal} {Living Reviews in Computational
  Astrophysics}\ }\textbf {\bibinfo {volume} {9}},\ \bibinfo {eid} {1}
  (\bibinfo {year} {2023})},\ \Eprint {http://arxiv.org/abs/2209.02538}
  {arXiv:2209.02538 [astro-ph.HE]} \BibitemShut {NoStop}%
\bibitem [{\citenamefont {Wanajo}\ \emph {et~al.}(2014)\citenamefont {Wanajo},
  \citenamefont {Sekiguchi}, \citenamefont {Nishimura}, \citenamefont {Kiuchi},
  \citenamefont {Kyutoku},\ and\ \citenamefont {Shibata}}]{Wanajo:2014wha}%
  \BibitemOpen
  \bibfield  {author} {\bibinfo {author} {\bibfnamefont {S.}~\bibnamefont
  {Wanajo}}, \bibinfo {author} {\bibfnamefont {Y.}~\bibnamefont {Sekiguchi}},
  \bibinfo {author} {\bibfnamefont {N.}~\bibnamefont {Nishimura}}, \bibinfo
  {author} {\bibfnamefont {K.}~\bibnamefont {Kiuchi}}, \bibinfo {author}
  {\bibfnamefont {K.}~\bibnamefont {Kyutoku}}, \ and\ \bibinfo {author}
  {\bibfnamefont {M.}~\bibnamefont {Shibata}},\ }\href {\doibase
  10.1088/2041-8205/789/2/L39} {\bibfield  {journal} {\bibinfo  {journal}
  {Astrophys. J. Lett.}\ }\textbf {\bibinfo {volume} {789}},\ \bibinfo {pages}
  {L39} (\bibinfo {year} {2014})},\ \Eprint {http://arxiv.org/abs/1402.7317}
  {arXiv:1402.7317 [astro-ph.SR]} \BibitemShut {NoStop}%
\bibitem [{\citenamefont {Martin}\ \emph {et~al.}(2018)\citenamefont {Martin},
  \citenamefont {Perego}, \citenamefont {Kastaun},\ and\ \citenamefont
  {Arcones}}]{Martin:2017dhc}%
  \BibitemOpen
  \bibfield  {author} {\bibinfo {author} {\bibfnamefont {D.}~\bibnamefont
  {Martin}}, \bibinfo {author} {\bibfnamefont {A.}~\bibnamefont {Perego}},
  \bibinfo {author} {\bibfnamefont {W.}~\bibnamefont {Kastaun}}, \ and\
  \bibinfo {author} {\bibfnamefont {A.}~\bibnamefont {Arcones}},\ }\href
  {\doibase 10.1088/1361-6382/aa9f5a} {\bibfield  {journal} {\bibinfo
  {journal} {Class. Quant. Grav.}\ }\textbf {\bibinfo {volume} {35}},\ \bibinfo
  {pages} {034001} (\bibinfo {year} {2018})},\ \Eprint
  {http://arxiv.org/abs/1710.04900} {arXiv:1710.04900 [astro-ph.HE]}
  \BibitemShut {NoStop}%
\bibitem [{\citenamefont {Radice}\ \emph {et~al.}(2018)\citenamefont {Radice},
  \citenamefont {Perego}, \citenamefont {Hotokezaka}, \citenamefont {Fromm},
  \citenamefont {Bernuzzi},\ and\ \citenamefont {Roberts}}]{Radice:2018pdn}%
  \BibitemOpen
  \bibfield  {author} {\bibinfo {author} {\bibfnamefont {D.}~\bibnamefont
  {Radice}}, \bibinfo {author} {\bibfnamefont {A.}~\bibnamefont {Perego}},
  \bibinfo {author} {\bibfnamefont {K.}~\bibnamefont {Hotokezaka}}, \bibinfo
  {author} {\bibfnamefont {S.~A.}\ \bibnamefont {Fromm}}, \bibinfo {author}
  {\bibfnamefont {S.}~\bibnamefont {Bernuzzi}}, \ and\ \bibinfo {author}
  {\bibfnamefont {L.~F.}\ \bibnamefont {Roberts}},\ }\href {\doibase
  10.3847/1538-4357/aaf054} {\bibfield  {journal} {\bibinfo  {journal}
  {Astrophys. J.}\ }\textbf {\bibinfo {volume} {869}},\ \bibinfo {pages} {130}
  (\bibinfo {year} {2018})},\ \Eprint {http://arxiv.org/abs/1809.11161}
  {arXiv:1809.11161 [astro-ph.HE]} \BibitemShut {NoStop}%
\bibitem [{\citenamefont {Kullmann}\ \emph {et~al.}(2022)\citenamefont
  {Kullmann}, \citenamefont {Goriely}, \citenamefont {Just}, \citenamefont
  {Ardevol-Pulpillo}, \citenamefont {Bauswein},\ and\ \citenamefont
  {Janka}}]{Kullmann:2021gvo}%
  \BibitemOpen
  \bibfield  {author} {\bibinfo {author} {\bibfnamefont {I.}~\bibnamefont
  {Kullmann}}, \bibinfo {author} {\bibfnamefont {S.}~\bibnamefont {Goriely}},
  \bibinfo {author} {\bibfnamefont {O.}~\bibnamefont {Just}}, \bibinfo {author}
  {\bibfnamefont {R.}~\bibnamefont {Ardevol-Pulpillo}}, \bibinfo {author}
  {\bibfnamefont {A.}~\bibnamefont {Bauswein}}, \ and\ \bibinfo {author}
  {\bibfnamefont {H.~T.}\ \bibnamefont {Janka}},\ }\href {\doibase
  10.1093/mnras/stab3393} {\bibfield  {journal} {\bibinfo  {journal} {Mon. Not.
  Roy. Astron. Soc.}\ }\textbf {\bibinfo {volume} {510}},\ \bibinfo {pages}
  {2804} (\bibinfo {year} {2022})},\ \Eprint {http://arxiv.org/abs/2109.02509}
  {arXiv:2109.02509 [astro-ph.HE]} \BibitemShut {NoStop}%
\bibitem [{\citenamefont {Fujibayashi}\ \emph {et~al.}(2023)\citenamefont
  {Fujibayashi}, \citenamefont {Kiuchi}, \citenamefont {Wanajo}, \citenamefont
  {Kyutoku}, \citenamefont {Sekiguchi},\ and\ \citenamefont
  {Shibata}}]{Fujibayashi:2022ftg}%
  \BibitemOpen
  \bibfield  {author} {\bibinfo {author} {\bibfnamefont {S.}~\bibnamefont
  {Fujibayashi}}, \bibinfo {author} {\bibfnamefont {K.}~\bibnamefont {Kiuchi}},
  \bibinfo {author} {\bibfnamefont {S.}~\bibnamefont {Wanajo}}, \bibinfo
  {author} {\bibfnamefont {K.}~\bibnamefont {Kyutoku}}, \bibinfo {author}
  {\bibfnamefont {Y.}~\bibnamefont {Sekiguchi}}, \ and\ \bibinfo {author}
  {\bibfnamefont {M.}~\bibnamefont {Shibata}},\ }\href {\doibase
  10.3847/1538-4357/ac9ce0} {\bibfield  {journal} {\bibinfo  {journal}
  {Astrophys. J.}\ }\textbf {\bibinfo {volume} {942}},\ \bibinfo {pages} {39}
  (\bibinfo {year} {2023})},\ \Eprint {http://arxiv.org/abs/2205.05557}
  {arXiv:2205.05557 [astro-ph.HE]} \BibitemShut {NoStop}%
\bibitem [{\citenamefont {Bruenn}(1985)}]{Bruenn:1985en}%
  \BibitemOpen
  \bibfield  {author} {\bibinfo {author} {\bibfnamefont {S.~W.}\ \bibnamefont
  {Bruenn}},\ }\href {\doibase 10.1086/191056} {\bibfield  {journal} {\bibinfo
  {journal} {Astrophys. J. Suppl.}\ }\textbf {\bibinfo {volume} {58}},\
  \bibinfo {pages} {771} (\bibinfo {year} {1985})}\BibitemShut {NoStop}%
\bibitem [{\citenamefont {Buras}\ \emph {et~al.}(2006)\citenamefont {Buras},
  \citenamefont {Rampp}, \citenamefont {Janka},\ and\ \citenamefont
  {Kifonidis}}]{Buras:2005rp}%
  \BibitemOpen
  \bibfield  {author} {\bibinfo {author} {\bibfnamefont {R.}~\bibnamefont
  {Buras}}, \bibinfo {author} {\bibfnamefont {M.}~\bibnamefont {Rampp}},
  \bibinfo {author} {\bibfnamefont {H.~T.}\ \bibnamefont {Janka}}, \ and\
  \bibinfo {author} {\bibfnamefont {K.}~\bibnamefont {Kifonidis}},\ }\href
  {\doibase 10.1051/0004-6361:20053783} {\bibfield  {journal} {\bibinfo
  {journal} {Astron. Astrophys.}\ }\textbf {\bibinfo {volume} {447}},\ \bibinfo
  {pages} {1049} (\bibinfo {year} {2006})},\ \Eprint
  {http://arxiv.org/abs/astro-ph/0507135} {arXiv:astro-ph/0507135} \BibitemShut
  {NoStop}%
\bibitem [{\citenamefont {Burrows}\ \emph {et~al.}(2006)\citenamefont
  {Burrows}, \citenamefont {Reddy},\ and\ \citenamefont
  {Thompson}}]{Burrows:2004vq}%
  \BibitemOpen
  \bibfield  {author} {\bibinfo {author} {\bibfnamefont {A.}~\bibnamefont
  {Burrows}}, \bibinfo {author} {\bibfnamefont {S.}~\bibnamefont {Reddy}}, \
  and\ \bibinfo {author} {\bibfnamefont {T.~A.}\ \bibnamefont {Thompson}},\
  }\href {\doibase 10.1016/j.nuclphysa.2004.06.012} {\bibfield  {journal}
  {\bibinfo  {journal} {Nucl. Phys. A}\ }\textbf {\bibinfo {volume} {777}},\
  \bibinfo {pages} {356} (\bibinfo {year} {2006})},\ \Eprint
  {http://arxiv.org/abs/astro-ph/0404432} {arXiv:astro-ph/0404432} \BibitemShut
  {NoStop}%
\bibitem [{\citenamefont {Pons}\ \emph {et~al.}(1999)\citenamefont {Pons},
  \citenamefont {Reddy}, \citenamefont {Prakash}, \citenamefont {Lattimer},\
  and\ \citenamefont {Miralles}}]{Pons:1998mm}%
  \BibitemOpen
  \bibfield  {author} {\bibinfo {author} {\bibfnamefont {J.~A.}\ \bibnamefont
  {Pons}}, \bibinfo {author} {\bibfnamefont {S.}~\bibnamefont {Reddy}},
  \bibinfo {author} {\bibfnamefont {M.}~\bibnamefont {Prakash}}, \bibinfo
  {author} {\bibfnamefont {J.~M.}\ \bibnamefont {Lattimer}}, \ and\ \bibinfo
  {author} {\bibfnamefont {J.~A.}\ \bibnamefont {Miralles}},\ }\href {\doibase
  10.1086/306889} {\bibfield  {journal} {\bibinfo  {journal} {Astrophys. J.}\
  }\textbf {\bibinfo {volume} {513}},\ \bibinfo {pages} {780} (\bibinfo {year}
  {1999})},\ \Eprint {http://arxiv.org/abs/astro-ph/9807040}
  {arXiv:astro-ph/9807040} \BibitemShut {NoStop}%
\bibitem [{\citenamefont {{Yakovlev}}\ \emph {et~al.}(2001)\citenamefont
  {{Yakovlev}}, \citenamefont {{Kaminker}}, \citenamefont {{Gnedin}},\ and\
  \citenamefont {{Haensel}}}]{Yakovlev2001}%
  \BibitemOpen
  \bibfield  {author} {\bibinfo {author} {\bibfnamefont {D.~G.}\ \bibnamefont
  {{Yakovlev}}}, \bibinfo {author} {\bibfnamefont {A.~D.}\ \bibnamefont
  {{Kaminker}}}, \bibinfo {author} {\bibfnamefont {O.~Y.}\ \bibnamefont
  {{Gnedin}}}, \ and\ \bibinfo {author} {\bibfnamefont {P.}~\bibnamefont
  {{Haensel}}},\ }\href {\doibase 10.1016/S0370-1573(00)00131-9} {\bibfield
  {journal} {\bibinfo  {journal} {Physics Reports}\ }\textbf {\bibinfo {volume}
  {354}},\ \bibinfo {pages} {1} (\bibinfo {year} {2001})},\ \Eprint
  {http://arxiv.org/abs/astro-ph/0012122} {arXiv:astro-ph/0012122 [astro-ph]}
  \BibitemShut {NoStop}%
\bibitem [{\citenamefont {{Page}}(2009)}]{Page2009}%
  \BibitemOpen
  \bibfield  {author} {\bibinfo {author} {\bibfnamefont {D.}~\bibnamefont
  {{Page}}},\ }in\ \href {\doibase 10.1007/978-3-540-76965-1_11} {\emph
  {\bibinfo {booktitle} {Astrophysics and Space Science Library}}},\ \bibinfo
  {series} {Astrophysics and Space Science Library}, Vol.\ \bibinfo {volume}
  {357},\ \bibinfo {editor} {edited by\ \bibinfo {editor} {\bibfnamefont
  {W.}~\bibnamefont {{Becker}}}}\ (\bibinfo {year} {2009})\ p.\ \bibinfo
  {pages} {247}\BibitemShut {NoStop}%
\bibitem [{\citenamefont {{Kiuchi}}\ \emph {et~al.}(2012)\citenamefont
  {{Kiuchi}}, \citenamefont {{Sekiguchi}}, \citenamefont {{Kyutoku}},\ and\
  \citenamefont {{Shibata}}}]{Kiuchi2012}%
  \BibitemOpen
  \bibfield  {author} {\bibinfo {author} {\bibfnamefont {K.}~\bibnamefont
  {{Kiuchi}}}, \bibinfo {author} {\bibfnamefont {Y.}~\bibnamefont
  {{Sekiguchi}}}, \bibinfo {author} {\bibfnamefont {K.}~\bibnamefont
  {{Kyutoku}}}, \ and\ \bibinfo {author} {\bibfnamefont {M.}~\bibnamefont
  {{Shibata}}},\ }in\ \href@noop {} {\emph {\bibinfo {booktitle} {Numerical
  Modeling of Space Plasma Slows (ASTRONUM 2011)}}},\ \bibinfo {series}
  {Astronomical Society of the Pacific Conference Series}, Vol.\ \bibinfo
  {volume} {459},\ \bibinfo {editor} {edited by\ \bibinfo {editor}
  {\bibfnamefont {N.~V.}\ \bibnamefont {{Pogorelov}}}, \bibinfo {editor}
  {\bibfnamefont {J.~A.}\ \bibnamefont {{Font}}}, \bibinfo {editor}
  {\bibfnamefont {E.}~\bibnamefont {{Audit}}}, \ and\ \bibinfo {editor}
  {\bibfnamefont {G.~P.}\ \bibnamefont {{Zank}}}}\ (\bibinfo {year} {2012})\
  p.~\bibinfo {pages} {85}\BibitemShut {NoStop}%
\bibitem [{\citenamefont {{Schmitt}}\ and\ \citenamefont
  {{Shternin}}(2018)}]{Schmitt2018}%
  \BibitemOpen
  \bibfield  {author} {\bibinfo {author} {\bibfnamefont {A.}~\bibnamefont
  {{Schmitt}}}\ and\ \bibinfo {author} {\bibfnamefont {P.}~\bibnamefont
  {{Shternin}}},\ }in\ \href {\doibase 10.1007/978-3-319-97616-7\_9} {\emph
  {\bibinfo {booktitle} {Astrophysics and Space Science Library}}},\ \bibinfo
  {series} {Astrophysics and Space Science Library}, Vol.\ \bibinfo {volume}
  {457},\ \bibinfo {editor} {edited by\ \bibinfo {editor} {\bibfnamefont
  {L.}~\bibnamefont {{Rezzolla}}}, \bibinfo {editor} {\bibfnamefont
  {P.}~\bibnamefont {{Pizzochero}}}, \bibinfo {editor} {\bibfnamefont {D.~I.}\
  \bibnamefont {{Jones}}}, \bibinfo {editor} {\bibfnamefont {N.}~\bibnamefont
  {{Rea}}}, \ and\ \bibinfo {editor} {\bibfnamefont {I.}~\bibnamefont
  {{Vida{\~n}a}}}}\ (\bibinfo {year} {2018})\ p.\ \bibinfo {pages} {455},\
  \Eprint {http://arxiv.org/abs/1711.06520} {arXiv:1711.06520 [astro-ph.HE]}
  \BibitemShut {NoStop}%
\bibitem [{\citenamefont {{Fujibayashi}}\ \emph {et~al.}(2020)\citenamefont
  {{Fujibayashi}}, \citenamefont {{Wanajo}}, \citenamefont {{Kiuchi}},
  \citenamefont {{Kyutoku}}, \citenamefont {{Sekiguchi}},\ and\ \citenamefont
  {{Shibata}}}]{Fujibayashi2020}%
  \BibitemOpen
  \bibfield  {author} {\bibinfo {author} {\bibfnamefont {S.}~\bibnamefont
  {{Fujibayashi}}}, \bibinfo {author} {\bibfnamefont {S.}~\bibnamefont
  {{Wanajo}}}, \bibinfo {author} {\bibfnamefont {K.}~\bibnamefont {{Kiuchi}}},
  \bibinfo {author} {\bibfnamefont {K.}~\bibnamefont {{Kyutoku}}}, \bibinfo
  {author} {\bibfnamefont {Y.}~\bibnamefont {{Sekiguchi}}}, \ and\ \bibinfo
  {author} {\bibfnamefont {M.}~\bibnamefont {{Shibata}}},\ }\href {\doibase
  10.3847/1538-4357/abafc2} {\bibfield  {journal} {\bibinfo  {journal} {\apj}\
  }\textbf {\bibinfo {volume} {901}},\ \bibinfo {eid} {122} (\bibinfo {year}
  {2020})},\ \Eprint {http://arxiv.org/abs/2007.00474} {arXiv:2007.00474
  [astro-ph.HE]} \BibitemShut {NoStop}%
\bibitem [{\citenamefont {{Friman}}\ and\ \citenamefont
  {{Maxwell}}(1979)}]{Friman1979}%
  \BibitemOpen
  \bibfield  {author} {\bibinfo {author} {\bibfnamefont {B.~L.}\ \bibnamefont
  {{Friman}}}\ and\ \bibinfo {author} {\bibfnamefont {O.~V.}\ \bibnamefont
  {{Maxwell}}},\ }\href {\doibase 10.1086/157313} {\bibfield  {journal}
  {\bibinfo  {journal} {The Astrophysical Journal}\ }\textbf {\bibinfo {volume}
  {232}},\ \bibinfo {pages} {541} (\bibinfo {year} {1979})}\BibitemShut
  {NoStop}%
\bibitem [{\citenamefont {{Lattimer}}\ \emph {et~al.}(1991)\citenamefont
  {{Lattimer}}, \citenamefont {{Pethick}}, \citenamefont {{Prakash}},\ and\
  \citenamefont {{Haensel}}}]{Lattimer1991}%
  \BibitemOpen
  \bibfield  {author} {\bibinfo {author} {\bibfnamefont {J.~M.}\ \bibnamefont
  {{Lattimer}}}, \bibinfo {author} {\bibfnamefont {C.~J.}\ \bibnamefont
  {{Pethick}}}, \bibinfo {author} {\bibfnamefont {M.}~\bibnamefont
  {{Prakash}}}, \ and\ \bibinfo {author} {\bibfnamefont {P.}~\bibnamefont
  {{Haensel}}},\ }\href {\doibase 10.1103/PhysRevLett.66.2701} {\bibfield
  {journal} {\bibinfo  {journal} {\prl}\ }\textbf {\bibinfo {volume} {66}},\
  \bibinfo {pages} {2701} (\bibinfo {year} {1991})}\BibitemShut {NoStop}%
\bibitem [{\citenamefont {{Blaschke}}\ \emph {et~al.}(1995)\citenamefont
  {{Blaschke}}, \citenamefont {{R{\"o}pke}}, \citenamefont {{Schulz}},
  \citenamefont {{Sedrakian}},\ and\ \citenamefont
  {{Voskresensky}}}]{Blaschke1995}%
  \BibitemOpen
  \bibfield  {author} {\bibinfo {author} {\bibfnamefont {D.}~\bibnamefont
  {{Blaschke}}}, \bibinfo {author} {\bibfnamefont {G.}~\bibnamefont
  {{R{\"o}pke}}}, \bibinfo {author} {\bibfnamefont {H.}~\bibnamefont
  {{Schulz}}}, \bibinfo {author} {\bibfnamefont {A.~D.}\ \bibnamefont
  {{Sedrakian}}}, \ and\ \bibinfo {author} {\bibfnamefont {D.~N.}\ \bibnamefont
  {{Voskresensky}}},\ }\href {\doibase 10.1093/mnras/273.3.596} {\bibfield
  {journal} {\bibinfo  {journal} {Monthly Notices of the Royal Astronomical
  Society}\ }\textbf {\bibinfo {volume} {273}},\ \bibinfo {pages} {596}
  (\bibinfo {year} {1995})}\BibitemShut {NoStop}%
\bibitem [{\citenamefont {{Yakovlev}}\ and\ \citenamefont
  {{Levenfish}}(1995)}]{Yakovlev1995}%
  \BibitemOpen
  \bibfield  {author} {\bibinfo {author} {\bibfnamefont {D.~G.}\ \bibnamefont
  {{Yakovlev}}}\ and\ \bibinfo {author} {\bibfnamefont {K.~P.}\ \bibnamefont
  {{Levenfish}}},\ }\href@noop {} {\bibfield  {journal} {\bibinfo  {journal}
  {Astronomy \& Astrophysics}\ }\textbf {\bibinfo {volume} {297}},\ \bibinfo
  {pages} {717} (\bibinfo {year} {1995})}\BibitemShut {NoStop}%
\bibitem [{\citenamefont {{Reddy}}\ \emph {et~al.}(1998)\citenamefont
  {{Reddy}}, \citenamefont {{Prakash}},\ and\ \citenamefont
  {{Lattimer}}}]{Reddy1998}%
  \BibitemOpen
  \bibfield  {author} {\bibinfo {author} {\bibfnamefont {S.}~\bibnamefont
  {{Reddy}}}, \bibinfo {author} {\bibfnamefont {M.}~\bibnamefont {{Prakash}}},
  \ and\ \bibinfo {author} {\bibfnamefont {J.~M.}\ \bibnamefont {{Lattimer}}},\
  }\href {\doibase 10.1103/PhysRevD.58.013009} {\bibfield  {journal} {\bibinfo
  {journal} {\prd}\ }\textbf {\bibinfo {volume} {58}},\ \bibinfo {eid} {013009}
  (\bibinfo {year} {1998})},\ \Eprint {http://arxiv.org/abs/astro-ph/9710115}
  {arXiv:astro-ph/9710115 [astro-ph]} \BibitemShut {NoStop}%
\bibitem [{\citenamefont {{Fu}}\ \emph {et~al.}(2008)\citenamefont {{Fu}},
  \citenamefont {{Wang}},\ and\ \citenamefont {{Liu}}}]{Fu2008}%
  \BibitemOpen
  \bibfield  {author} {\bibinfo {author} {\bibfnamefont {W.-j.}\ \bibnamefont
  {{Fu}}}, \bibinfo {author} {\bibfnamefont {G.-h.}\ \bibnamefont {{Wang}}}, \
  and\ \bibinfo {author} {\bibfnamefont {Y.-x.}\ \bibnamefont {{Liu}}},\ }\href
  {\doibase 10.1086/528361} {\bibfield  {journal} {\bibinfo  {journal} {The
  Astrophysical Journal}\ }\textbf {\bibinfo {volume} {678}},\ \bibinfo {pages}
  {1517} (\bibinfo {year} {2008})}\BibitemShut {NoStop}%
\bibitem [{\citenamefont {{Roberts}}\ and\ \citenamefont
  {{Reddy}}(2017)}]{Roberts2017}%
  \BibitemOpen
  \bibfield  {author} {\bibinfo {author} {\bibfnamefont {L.~F.}\ \bibnamefont
  {{Roberts}}}\ and\ \bibinfo {author} {\bibfnamefont {S.}~\bibnamefont
  {{Reddy}}},\ }\href {\doibase 10.1103/PhysRevC.95.045807} {\bibfield
  {journal} {\bibinfo  {journal} {Physical Review C}\ }\textbf {\bibinfo
  {volume} {95}},\ \bibinfo {eid} {045807} (\bibinfo {year} {2017})},\ \Eprint
  {http://arxiv.org/abs/1612.02764} {arXiv:1612.02764 [astro-ph.HE]}
  \BibitemShut {NoStop}%
\bibitem [{\citenamefont {{Alford}}\ and\ \citenamefont
  {{Harris}}(2018)}]{Alford2018}%
  \BibitemOpen
  \bibfield  {author} {\bibinfo {author} {\bibfnamefont {M.~G.}\ \bibnamefont
  {{Alford}}}\ and\ \bibinfo {author} {\bibfnamefont {S.~P.}\ \bibnamefont
  {{Harris}}},\ }\href {\doibase 10.1103/PhysRevC.98.065806} {\bibfield
  {journal} {\bibinfo  {journal} {Physical Review C}\ }\textbf {\bibinfo
  {volume} {98}},\ \bibinfo {eid} {065806} (\bibinfo {year} {2018})},\ \Eprint
  {http://arxiv.org/abs/1803.00662} {arXiv:1803.00662 [nucl-th]} \BibitemShut
  {NoStop}%
\bibitem [{\citenamefont {{Fischer}}\ \emph {et~al.}(2020)\citenamefont
  {{Fischer}}, \citenamefont {{Guo}}, \citenamefont {{Dzhioev}}, \citenamefont
  {{Mart{\'\i}nez-Pinedo}}, \citenamefont {{Wu}}, \citenamefont {{Lohs}},\ and\
  \citenamefont {{Qian}}}]{Fischer2020}%
  \BibitemOpen
  \bibfield  {author} {\bibinfo {author} {\bibfnamefont {T.}~\bibnamefont
  {{Fischer}}}, \bibinfo {author} {\bibfnamefont {G.}~\bibnamefont {{Guo}}},
  \bibinfo {author} {\bibfnamefont {A.~A.}\ \bibnamefont {{Dzhioev}}}, \bibinfo
  {author} {\bibfnamefont {G.}~\bibnamefont {{Mart{\'\i}nez-Pinedo}}}, \bibinfo
  {author} {\bibfnamefont {M.-R.}\ \bibnamefont {{Wu}}}, \bibinfo {author}
  {\bibfnamefont {A.}~\bibnamefont {{Lohs}}}, \ and\ \bibinfo {author}
  {\bibfnamefont {Y.-Z.}\ \bibnamefont {{Qian}}},\ }\href {\doibase
  10.1103/PhysRevC.101.025804} {\bibfield  {journal} {\bibinfo  {journal}
  {Physical Review C}\ }\textbf {\bibinfo {volume} {101}},\ \bibinfo {eid}
  {025804} (\bibinfo {year} {2020})},\ \Eprint
  {http://arxiv.org/abs/1804.10890} {arXiv:1804.10890 [astro-ph.HE]}
  \BibitemShut {NoStop}%
\bibitem [{\citenamefont {{Oertel}}\ \emph {et~al.}(2020)\citenamefont
  {{Oertel}}, \citenamefont {{Pascal}}, \citenamefont {{Mancini}},\ and\
  \citenamefont {{Novak}}}]{Oertel2020}%
  \BibitemOpen
  \bibfield  {author} {\bibinfo {author} {\bibfnamefont {M.}~\bibnamefont
  {{Oertel}}}, \bibinfo {author} {\bibfnamefont {A.}~\bibnamefont {{Pascal}}},
  \bibinfo {author} {\bibfnamefont {M.}~\bibnamefont {{Mancini}}}, \ and\
  \bibinfo {author} {\bibfnamefont {J.}~\bibnamefont {{Novak}}},\ }\href
  {\doibase 10.1103/PhysRevC.102.035802} {\bibfield  {journal} {\bibinfo
  {journal} {Physical Review C}\ }\textbf {\bibinfo {volume} {102}},\ \bibinfo
  {eid} {035802} (\bibinfo {year} {2020})},\ \Eprint
  {http://arxiv.org/abs/2003.02152} {arXiv:2003.02152 [astro-ph.HE]}
  \BibitemShut {NoStop}%
\bibitem [{\citenamefont {Shin}\ \emph {et~al.}(2023)\citenamefont {Shin},
  \citenamefont {Rrapaj}, \citenamefont {Holt},\ and\ \citenamefont
  {Reddy}}]{Shin:2023sei}%
  \BibitemOpen
  \bibfield  {author} {\bibinfo {author} {\bibfnamefont {E.}~\bibnamefont
  {Shin}}, \bibinfo {author} {\bibfnamefont {E.}~\bibnamefont {Rrapaj}},
  \bibinfo {author} {\bibfnamefont {J.~W.}\ \bibnamefont {Holt}}, \ and\
  \bibinfo {author} {\bibfnamefont {S.~K.}\ \bibnamefont {Reddy}},\ }\href@noop
  {} {\  (\bibinfo {year} {2023})},\ \Eprint {http://arxiv.org/abs/2306.05280}
  {arXiv:2306.05280 [nucl-th]} \BibitemShut {NoStop}%
\bibitem [{\citenamefont {{Chiu}}\ and\ \citenamefont
  {{Salpeter}}(1964)}]{Chiu1964}%
  \BibitemOpen
  \bibfield  {author} {\bibinfo {author} {\bibfnamefont {H.~Y.}\ \bibnamefont
  {{Chiu}}}\ and\ \bibinfo {author} {\bibfnamefont {E.~E.}\ \bibnamefont
  {{Salpeter}}},\ }\href {\doibase 10.1103/PhysRevLett.12.413} {\bibfield
  {journal} {\bibinfo  {journal} {Physical Review Letters}\ }\textbf {\bibinfo
  {volume} {12}},\ \bibinfo {pages} {413} (\bibinfo {year} {1964})}\BibitemShut
  {NoStop}%
\bibitem [{\citenamefont {{Shternin}}\ \emph {et~al.}(2018)\citenamefont
  {{Shternin}}, \citenamefont {{Baldo}},\ and\ \citenamefont
  {{Haensel}}}]{Shternin2018}%
  \BibitemOpen
  \bibfield  {author} {\bibinfo {author} {\bibfnamefont {P.~S.}\ \bibnamefont
  {{Shternin}}}, \bibinfo {author} {\bibfnamefont {M.}~\bibnamefont {{Baldo}}},
  \ and\ \bibinfo {author} {\bibfnamefont {P.}~\bibnamefont {{Haensel}}},\
  }\href {\doibase 10.1016/j.physletb.2018.09.035} {\bibfield  {journal}
  {\bibinfo  {journal} {Physics Letters B}\ }\textbf {\bibinfo {volume}
  {786}},\ \bibinfo {pages} {28} (\bibinfo {year} {2018})},\ \Eprint
  {http://arxiv.org/abs/1807.06569} {arXiv:1807.06569 [astro-ph.HE]}
  \BibitemShut {NoStop}%
\bibitem [{\citenamefont {Sigl}(1997)}]{Sigl:1997ga}%
  \BibitemOpen
  \bibfield  {author} {\bibinfo {author} {\bibfnamefont {G.}~\bibnamefont
  {Sigl}},\ }\href {\doibase 10.1103/PhysRevD.56.3179} {\bibfield  {journal}
  {\bibinfo  {journal} {Phys. Rev. D}\ }\textbf {\bibinfo {volume} {56}},\
  \bibinfo {pages} {3179} (\bibinfo {year} {1997})},\ \Eprint
  {http://arxiv.org/abs/astro-ph/9703056} {arXiv:astro-ph/9703056} \BibitemShut
  {NoStop}%
\bibitem [{\citenamefont {{Hannestad}}\ and\ \citenamefont
  {{Raffelt}}(1998)}]{Hannestad1998}%
  \BibitemOpen
  \bibfield  {author} {\bibinfo {author} {\bibfnamefont {S.}~\bibnamefont
  {{Hannestad}}}\ and\ \bibinfo {author} {\bibfnamefont {G.}~\bibnamefont
  {{Raffelt}}},\ }\href {\doibase 10.1086/306303} {\bibfield  {journal}
  {\bibinfo  {journal} {The Astrophysical Journal}\ }\textbf {\bibinfo {volume}
  {507}},\ \bibinfo {pages} {339} (\bibinfo {year} {1998})},\ \Eprint
  {http://arxiv.org/abs/astro-ph/9711132} {arXiv:astro-ph/9711132 [astro-ph]}
  \BibitemShut {NoStop}%
\bibitem [{\citenamefont {{Sedrakian}}\ and\ \citenamefont
  {{Dieperink}}(2000)}]{Sedrakian2000}%
  \BibitemOpen
  \bibfield  {author} {\bibinfo {author} {\bibfnamefont {A.}~\bibnamefont
  {{Sedrakian}}}\ and\ \bibinfo {author} {\bibfnamefont {A.~E.~L.}\
  \bibnamefont {{Dieperink}}},\ }\href {\doibase 10.1103/PhysRevD.62.083002}
  {\bibfield  {journal} {\bibinfo  {journal} {\prd}\ }\textbf {\bibinfo
  {volume} {62}},\ \bibinfo {eid} {083002} (\bibinfo {year} {2000})},\ \Eprint
  {http://arxiv.org/abs/astro-ph/0002228} {arXiv:astro-ph/0002228 [astro-ph]}
  \BibitemShut {NoStop}%
\bibitem [{\citenamefont {Hanhart}\ \emph {et~al.}(2001)\citenamefont
  {Hanhart}, \citenamefont {Phillips},\ and\ \citenamefont
  {Reddy}}]{Hanhart:2000ae}%
  \BibitemOpen
  \bibfield  {author} {\bibinfo {author} {\bibfnamefont {C.}~\bibnamefont
  {Hanhart}}, \bibinfo {author} {\bibfnamefont {D.~R.}\ \bibnamefont
  {Phillips}}, \ and\ \bibinfo {author} {\bibfnamefont {S.}~\bibnamefont
  {Reddy}},\ }\href {\doibase 10.1016/S0370-2693(00)01382-4} {\bibfield
  {journal} {\bibinfo  {journal} {Phys. Lett. B}\ }\textbf {\bibinfo {volume}
  {499}},\ \bibinfo {pages} {9} (\bibinfo {year} {2001})},\ \Eprint
  {http://arxiv.org/abs/astro-ph/0003445} {arXiv:astro-ph/0003445} \BibitemShut
  {NoStop}%
\bibitem [{\citenamefont {Guo}\ and\ \citenamefont
  {Mart\'\i{}nez-Pinedo}(2019)}]{Guo:2019cvs}%
  \BibitemOpen
  \bibfield  {author} {\bibinfo {author} {\bibfnamefont {G.}~\bibnamefont
  {Guo}}\ and\ \bibinfo {author} {\bibfnamefont {G.}~\bibnamefont
  {Mart\'\i{}nez-Pinedo}},\ }\href {\doibase 10.3847/1538-4357/ab536d}
  {\bibfield  {journal} {\bibinfo  {journal} {Astrophys. J.}\ }\textbf
  {\bibinfo {volume} {887}},\ \bibinfo {pages} {58} (\bibinfo {year} {2019})},\
  \Eprint {http://arxiv.org/abs/1905.13634} {arXiv:1905.13634 [astro-ph.HE]}
  \BibitemShut {NoStop}%
\bibitem [{\citenamefont {Alford}\ \emph {et~al.}(2021)\citenamefont {Alford},
  \citenamefont {Haber}, \citenamefont {Harris},\ and\ \citenamefont
  {Zhang}}]{Alford:2021ogv}%
  \BibitemOpen
  \bibfield  {author} {\bibinfo {author} {\bibfnamefont {M.~G.}\ \bibnamefont
  {Alford}}, \bibinfo {author} {\bibfnamefont {A.}~\bibnamefont {Haber}},
  \bibinfo {author} {\bibfnamefont {S.~P.}\ \bibnamefont {Harris}}, \ and\
  \bibinfo {author} {\bibfnamefont {Z.}~\bibnamefont {Zhang}},\ }\href
  {\doibase 10.3390/universe7110399} {\bibfield  {journal} {\bibinfo  {journal}
  {Universe}\ }\textbf {\bibinfo {volume} {7}},\ \bibinfo {pages} {399}
  (\bibinfo {year} {2021})},\ \Eprint {http://arxiv.org/abs/2108.03324}
  {arXiv:2108.03324 [nucl-th]} \BibitemShut {NoStop}%
\bibitem [{\citenamefont {Roberts}\ \emph {et~al.}(2012)\citenamefont
  {Roberts}, \citenamefont {Reddy},\ and\ \citenamefont
  {Shen}}]{Roberts:2012um}%
  \BibitemOpen
  \bibfield  {author} {\bibinfo {author} {\bibfnamefont {L.~F.}\ \bibnamefont
  {Roberts}}, \bibinfo {author} {\bibfnamefont {S.}~\bibnamefont {Reddy}}, \
  and\ \bibinfo {author} {\bibfnamefont {G.}~\bibnamefont {Shen}},\ }\href
  {\doibase 10.1103/PhysRevC.86.065803} {\bibfield  {journal} {\bibinfo
  {journal} {Phys. Rev. C}\ }\textbf {\bibinfo {volume} {86}},\ \bibinfo
  {pages} {065803} (\bibinfo {year} {2012})},\ \Eprint
  {http://arxiv.org/abs/1205.4066} {arXiv:1205.4066 [astro-ph.HE]} \BibitemShut
  {NoStop}%
\bibitem [{\citenamefont {{Schneider}}\ \emph {et~al.}(2019)\citenamefont
  {{Schneider}}, \citenamefont {{Constantinou}}, \citenamefont {{Muccioli}},\
  and\ \citenamefont {{Prakash}}}]{Schneider2019}%
  \BibitemOpen
  \bibfield  {author} {\bibinfo {author} {\bibfnamefont {A.~S.}\ \bibnamefont
  {{Schneider}}}, \bibinfo {author} {\bibfnamefont {C.}~\bibnamefont
  {{Constantinou}}}, \bibinfo {author} {\bibfnamefont {B.}~\bibnamefont
  {{Muccioli}}}, \ and\ \bibinfo {author} {\bibfnamefont {M.}~\bibnamefont
  {{Prakash}}},\ }\href {\doibase 10.1103/PhysRevC.100.025803} {\bibfield
  {journal} {\bibinfo  {journal} {Physical Review C}\ }\textbf {\bibinfo
  {volume} {100}},\ \bibinfo {eid} {025803} (\bibinfo {year} {2019})},\ \Eprint
  {http://arxiv.org/abs/1901.09652} {arXiv:1901.09652 [nucl-th]} \BibitemShut
  {NoStop}%
\bibitem [{\citenamefont {{CompOSE}}(2023)}]{aprCompose}%
  \BibitemOpen
  \bibfield  {author} {\bibinfo {author} {\bibnamefont {{CompOSE}}},\ }\href
  {https://compose.obspm.fr/eos/149} {\enquote {\bibinfo {title} {Sro(apr) sna
  version general purpose equation of state as of march 7, 2023.}}\ } (\bibinfo
  {year} {2023})\BibitemShut {NoStop}%
\bibitem [{\citenamefont {Typel}\ \emph {et~al.}(2015)\citenamefont {Typel},
  \citenamefont {Oertel},\ and\ \citenamefont {Kl\"ahn}}]{Typel:2013rza}%
  \BibitemOpen
  \bibfield  {author} {\bibinfo {author} {\bibfnamefont {S.}~\bibnamefont
  {Typel}}, \bibinfo {author} {\bibfnamefont {M.}~\bibnamefont {Oertel}}, \
  and\ \bibinfo {author} {\bibfnamefont {T.}~\bibnamefont {Kl\"ahn}},\ }\href
  {\doibase 10.1134/S1063779615040061} {\bibfield  {journal} {\bibinfo
  {journal} {Phys. Part. Nucl.}\ }\textbf {\bibinfo {volume} {46}},\ \bibinfo
  {pages} {633} (\bibinfo {year} {2015})},\ \Eprint
  {http://arxiv.org/abs/1307.5715} {arXiv:1307.5715 [astro-ph.SR]} \BibitemShut
  {NoStop}%
\bibitem [{\citenamefont {Typel}\ \emph {et~al.}(2022)\citenamefont {Typel}
  \emph {et~al.}}]{CompOSECoreTeam:2022ddl}%
  \BibitemOpen
  \bibfield  {author} {\bibinfo {author} {\bibfnamefont {S.}~\bibnamefont
  {Typel}} \emph {et~al.} (\bibinfo {collaboration} {CompOSE Core Team}),\
  }\href {\doibase 10.1140/epja/s10050-022-00847-y} {\bibfield  {journal}
  {\bibinfo  {journal} {Eur. Phys. J. A}\ }\textbf {\bibinfo {volume} {58}},\
  \bibinfo {pages} {221} (\bibinfo {year} {2022})},\ \Eprint
  {http://arxiv.org/abs/2203.03209} {arXiv:2203.03209 [astro-ph.HE]}
  \BibitemShut {NoStop}%
\bibitem [{\citenamefont {Sedrakian}\ and\ \citenamefont
  {Dieperink}(1999)}]{Sedrakian:1999jh}%
  \BibitemOpen
  \bibfield  {author} {\bibinfo {author} {\bibfnamefont {A.}~\bibnamefont
  {Sedrakian}}\ and\ \bibinfo {author} {\bibfnamefont {A.}~\bibnamefont
  {Dieperink}},\ }\href {\doibase 10.1016/S0370-2693(99)00989-2} {\bibfield
  {journal} {\bibinfo  {journal} {Phys. Lett. B}\ }\textbf {\bibinfo {volume}
  {463}},\ \bibinfo {pages} {145} (\bibinfo {year} {1999})},\ \Eprint
  {http://arxiv.org/abs/nucl-th/9905039} {arXiv:nucl-th/9905039} \BibitemShut
  {NoStop}%
\bibitem [{\citenamefont {Sedrakian}(2007)}]{Sedrakian:2006mq}%
  \BibitemOpen
  \bibfield  {author} {\bibinfo {author} {\bibfnamefont {A.}~\bibnamefont
  {Sedrakian}},\ }\href {\doibase 10.1016/j.ppnp.2006.02.002} {\bibfield
  {journal} {\bibinfo  {journal} {Prog. Part. Nucl. Phys.}\ }\textbf {\bibinfo
  {volume} {58}},\ \bibinfo {pages} {168} (\bibinfo {year} {2007})},\ \Eprint
  {http://arxiv.org/abs/nucl-th/0601086} {arXiv:nucl-th/0601086} \BibitemShut
  {NoStop}%
\bibitem [{\citenamefont {{Schmitt}}\ \emph {et~al.}(2006)\citenamefont
  {{Schmitt}}, \citenamefont {{Shovkovy}},\ and\ \citenamefont
  {{Wang}}}]{Schmitt2006}%
  \BibitemOpen
  \bibfield  {author} {\bibinfo {author} {\bibfnamefont {A.}~\bibnamefont
  {{Schmitt}}}, \bibinfo {author} {\bibfnamefont {I.~A.}\ \bibnamefont
  {{Shovkovy}}}, \ and\ \bibinfo {author} {\bibfnamefont {Q.}~\bibnamefont
  {{Wang}}},\ }\href {\doibase 10.1103/PhysRevD.73.034012} {\bibfield
  {journal} {\bibinfo  {journal} {\prd}\ }\textbf {\bibinfo {volume} {73}},\
  \bibinfo {eid} {034012} (\bibinfo {year} {2006})},\ \Eprint
  {http://arxiv.org/abs/hep-ph/0510347} {arXiv:hep-ph/0510347 [hep-ph]}
  \BibitemShut {NoStop}%
\bibitem [{\citenamefont {{Bacca}}\ \emph {et~al.}(2012)\citenamefont
  {{Bacca}}, \citenamefont {{Hally}}, \citenamefont {{Liebend{\"o}rfer}},
  \citenamefont {{Perego}}, \citenamefont {{Pethick}},\ and\ \citenamefont
  {{Schwenk}}}]{Bacca2012}%
  \BibitemOpen
  \bibfield  {author} {\bibinfo {author} {\bibfnamefont {S.}~\bibnamefont
  {{Bacca}}}, \bibinfo {author} {\bibfnamefont {K.}~\bibnamefont {{Hally}}},
  \bibinfo {author} {\bibfnamefont {M.}~\bibnamefont {{Liebend{\"o}rfer}}},
  \bibinfo {author} {\bibfnamefont {A.}~\bibnamefont {{Perego}}}, \bibinfo
  {author} {\bibfnamefont {C.~J.}\ \bibnamefont {{Pethick}}}, \ and\ \bibinfo
  {author} {\bibfnamefont {A.}~\bibnamefont {{Schwenk}}},\ }\href {\doibase
  10.1088/0004-637X/758/1/34} {\bibfield  {journal} {\bibinfo  {journal} {The
  Astrophysical Journal}\ }\textbf {\bibinfo {volume} {758}},\ \bibinfo {eid}
  {34} (\bibinfo {year} {2012})},\ \Eprint {http://arxiv.org/abs/1112.5185}
  {arXiv:1112.5185 [astro-ph.HE]} \BibitemShut {NoStop}%
\bibitem [{\citenamefont {Leinson}\ and\ \citenamefont
  {Perez}(2001)}]{Leinson:2001ei}%
  \BibitemOpen
  \bibfield  {author} {\bibinfo {author} {\bibfnamefont {L.~B.}\ \bibnamefont
  {Leinson}}\ and\ \bibinfo {author} {\bibfnamefont {A.}~\bibnamefont
  {Perez}},\ }\href {\doibase 10.1016/S0370-2693(01)01042-5} {\bibfield
  {journal} {\bibinfo  {journal} {Phys. Lett. B}\ }\textbf {\bibinfo {volume}
  {518}},\ \bibinfo {pages} {15} (\bibinfo {year} {2001})},\ \bibinfo {note}
  {[Erratum: Phys.Lett.B 522, 358--358 (2001)]},\ \Eprint
  {http://arxiv.org/abs/hep-ph/0110207} {arXiv:hep-ph/0110207} \BibitemShut
  {NoStop}%
\bibitem [{\citenamefont {Leinson}(2002)}]{Leinson:2002bw}%
  \BibitemOpen
  \bibfield  {author} {\bibinfo {author} {\bibfnamefont {L.~B.}\ \bibnamefont
  {Leinson}},\ }\href {\doibase 10.1016/S0375-9474(02)00991-0} {\bibfield
  {journal} {\bibinfo  {journal} {Nucl. Phys. A}\ }\textbf {\bibinfo {volume}
  {707}},\ \bibinfo {pages} {543} (\bibinfo {year} {2002})},\ \Eprint
  {http://arxiv.org/abs/hep-ph/0207116} {arXiv:hep-ph/0207116} \BibitemShut
  {NoStop}%
\bibitem [{\citenamefont {{Alford}}\ \emph {et~al.}(2022)\citenamefont
  {{Alford}}, \citenamefont {{Harutyunyan}},\ and\ \citenamefont
  {{Sedrakian}}}]{Alford2022}%
  \BibitemOpen
  \bibfield  {author} {\bibinfo {author} {\bibfnamefont {M.}~\bibnamefont
  {{Alford}}}, \bibinfo {author} {\bibfnamefont {A.}~\bibnamefont
  {{Harutyunyan}}}, \ and\ \bibinfo {author} {\bibfnamefont {A.}~\bibnamefont
  {{Sedrakian}}},\ }\href {\doibase 10.3390/particles5030029} {\bibfield
  {journal} {\bibinfo  {journal} {Particles}\ }\textbf {\bibinfo {volume}
  {5}},\ \bibinfo {pages} {361} (\bibinfo {year} {2022})},\ \Eprint
  {http://arxiv.org/abs/2209.04717} {arXiv:2209.04717 [astro-ph.HE]}
  \BibitemShut {NoStop}%
\bibitem [{\citenamefont {Rapp}\ \emph {et~al.}(1998)\citenamefont {Rapp},
  \citenamefont {Urban}, \citenamefont {Buballa},\ and\ \citenamefont
  {Wambach}}]{Rapp:1997ei}%
  \BibitemOpen
  \bibfield  {author} {\bibinfo {author} {\bibfnamefont {R.}~\bibnamefont
  {Rapp}}, \bibinfo {author} {\bibfnamefont {M.}~\bibnamefont {Urban}},
  \bibinfo {author} {\bibfnamefont {M.}~\bibnamefont {Buballa}}, \ and\
  \bibinfo {author} {\bibfnamefont {J.}~\bibnamefont {Wambach}},\ }\href
  {\doibase 10.1016/S0370-2693(97)01360-9} {\bibfield  {journal} {\bibinfo
  {journal} {Phys. Lett. B}\ }\textbf {\bibinfo {volume} {417}},\ \bibinfo
  {pages} {1} (\bibinfo {year} {1998})},\ \Eprint
  {http://arxiv.org/abs/nucl-th/9709008} {arXiv:nucl-th/9709008} \BibitemShut
  {NoStop}%
\bibitem [{\citenamefont {{Kaminker}}\ \emph {et~al.}(2016)\citenamefont
  {{Kaminker}}, \citenamefont {{Yakovlev}},\ and\ \citenamefont
  {{Haensel}}}]{Kaminker2016}%
  \BibitemOpen
  \bibfield  {author} {\bibinfo {author} {\bibfnamefont {A.~D.}\ \bibnamefont
  {{Kaminker}}}, \bibinfo {author} {\bibfnamefont {D.~G.}\ \bibnamefont
  {{Yakovlev}}}, \ and\ \bibinfo {author} {\bibfnamefont {P.}~\bibnamefont
  {{Haensel}}},\ }\href {\doibase 10.1007/s10509-016-2854-5} {\bibfield
  {journal} {\bibinfo  {journal} {Astrophysics and Space Science}\ }\textbf
  {\bibinfo {volume} {361}},\ \bibinfo {eid} {267} (\bibinfo {year} {2016})},\
  \Eprint {http://arxiv.org/abs/1607.05265} {arXiv:1607.05265 [astro-ph.HE]}
  \BibitemShut {NoStop}%
\bibitem [{\citenamefont {Martinez-Pinedo}\ \emph {et~al.}(2012)\citenamefont
  {Martinez-Pinedo}, \citenamefont {Fischer}, \citenamefont {Lohs},\ and\
  \citenamefont {H{\"u}ther}}]{Martinez-Pinedo:2012eaj}%
  \BibitemOpen
  \bibfield  {author} {\bibinfo {author} {\bibfnamefont {G.}~\bibnamefont
  {Martinez-Pinedo}}, \bibinfo {author} {\bibfnamefont {T.}~\bibnamefont
  {Fischer}}, \bibinfo {author} {\bibfnamefont {A.}~\bibnamefont {Lohs}}, \
  and\ \bibinfo {author} {\bibfnamefont {L.}~\bibnamefont {H{\"u}ther}},\
  }\href {\doibase 10.1103/PhysRevLett.109.251104} {\bibfield  {journal}
  {\bibinfo  {journal} {Phys. Rev. Lett.}\ }\textbf {\bibinfo {volume} {109}},\
  \bibinfo {pages} {251104} (\bibinfo {year} {2012})},\ \Eprint
  {http://arxiv.org/abs/1205.2793} {arXiv:1205.2793 [astro-ph.HE]} \BibitemShut
  {NoStop}%
\bibitem [{\citenamefont {Lykasov}\ \emph {et~al.}(2008)\citenamefont
  {Lykasov}, \citenamefont {Pethick},\ and\ \citenamefont
  {Schwenk}}]{Lykasov:2008yz}%
  \BibitemOpen
  \bibfield  {author} {\bibinfo {author} {\bibfnamefont {G.~I.}\ \bibnamefont
  {Lykasov}}, \bibinfo {author} {\bibfnamefont {C.~J.}\ \bibnamefont
  {Pethick}}, \ and\ \bibinfo {author} {\bibfnamefont {A.}~\bibnamefont
  {Schwenk}},\ }\href {\doibase 10.1103/PhysRevC.78.045803} {\bibfield
  {journal} {\bibinfo  {journal} {Phys. Rev. C}\ }\textbf {\bibinfo {volume}
  {78}},\ \bibinfo {pages} {045803} (\bibinfo {year} {2008})},\ \Eprint
  {http://arxiv.org/abs/0808.0330} {arXiv:0808.0330 [nucl-th]} \BibitemShut
  {NoStop}%
\bibitem [{\citenamefont {{Bartl}}(2016)}]{Bartl2016}%
  \BibitemOpen
  \bibfield  {author} {\bibinfo {author} {\bibfnamefont {A.}~\bibnamefont
  {{Bartl}}},\ }\emph {\bibinfo {title} {{Neutrino Interactions with Supernova
  Matter}}},\ \href@noop {} {Ph.D. thesis},\ \bibinfo  {school} {Technical
  University of Darmstadt, Germany} (\bibinfo {year} {2016})\BibitemShut
  {NoStop}%
\bibitem [{\citenamefont {Voskresensky}\ and\ \citenamefont
  {Senatorov}(1987)}]{Voskresensky:1987hm}%
  \BibitemOpen
  \bibfield  {author} {\bibinfo {author} {\bibfnamefont {D.~N.}\ \bibnamefont
  {Voskresensky}}\ and\ \bibinfo {author} {\bibfnamefont {A.~V.}\ \bibnamefont
  {Senatorov}},\ }\href@noop {} {\bibfield  {journal} {\bibinfo  {journal}
  {Sov. J. Nucl. Phys.}\ }\textbf {\bibinfo {volume} {45}},\ \bibinfo {pages}
  {411} (\bibinfo {year} {1987})}\BibitemShut {NoStop}%
\bibitem [{\citenamefont {Knoll}\ and\ \citenamefont
  {Voskresensky}(1996)}]{Knoll:1995nz}%
  \BibitemOpen
  \bibfield  {author} {\bibinfo {author} {\bibfnamefont {J.}~\bibnamefont
  {Knoll}}\ and\ \bibinfo {author} {\bibfnamefont {D.~N.}\ \bibnamefont
  {Voskresensky}},\ }\href {\doibase 10.1006/aphy.1996.0082} {\bibfield
  {journal} {\bibinfo  {journal} {Annals Phys.}\ }\textbf {\bibinfo {volume}
  {249}},\ \bibinfo {pages} {532} (\bibinfo {year} {1996})},\ \Eprint
  {http://arxiv.org/abs/hep-ph/9510417} {arXiv:hep-ph/9510417} \BibitemShut
  {NoStop}%
\bibitem [{\citenamefont {Matsubara}(1955)}]{Matsubara1955}%
  \BibitemOpen
  \bibfield  {author} {\bibinfo {author} {\bibfnamefont {T.}~\bibnamefont
  {Matsubara}},\ }\href {\doibase 10.1143/PTP.14.351} {\bibfield  {journal}
  {\bibinfo  {journal} {Progress of Theoretical Physics}\ }\textbf {\bibinfo
  {volume} {14}},\ \bibinfo {pages} {351} (\bibinfo {year} {1955})},\ \Eprint
  {http://arxiv.org/abs/https://academic.oup.com/ptp/article-pdf/14/4/351/5286981/14-4-351.pdf}
  {https://academic.oup.com/ptp/article-pdf/14/4/351/5286981/14-4-351.pdf}
  \BibitemShut {NoStop}%
\bibitem [{\citenamefont {{Metropolis}}\ \emph {et~al.}(1953)\citenamefont
  {{Metropolis}}, \citenamefont {{Rosenbluth}}, \citenamefont {{Rosenbluth}},
  \citenamefont {{Teller}},\ and\ \citenamefont {{Teller}}}]{Metropolis1953}%
  \BibitemOpen
  \bibfield  {author} {\bibinfo {author} {\bibfnamefont {N.}~\bibnamefont
  {{Metropolis}}}, \bibinfo {author} {\bibfnamefont {A.~W.}\ \bibnamefont
  {{Rosenbluth}}}, \bibinfo {author} {\bibfnamefont {M.~N.}\ \bibnamefont
  {{Rosenbluth}}}, \bibinfo {author} {\bibfnamefont {A.~H.}\ \bibnamefont
  {{Teller}}}, \ and\ \bibinfo {author} {\bibfnamefont {E.}~\bibnamefont
  {{Teller}}},\ }\href {\doibase 10.1063/1.1699114} {\bibfield  {journal}
  {\bibinfo  {journal} {Journal of Computational Physics}\ }\textbf {\bibinfo
  {volume} {21}},\ \bibinfo {pages} {1087} (\bibinfo {year}
  {1953})}\BibitemShut {NoStop}%
\end{thebibliography}
\end{document}